\documentclass[usenatbib]{mnras}
\pdfoutput=1

\usepackage[T1]{fontenc}

\DeclareRobustCommand{\VAN}[3]{#2}
\let\VANthebibliography\thebibliography
\def\thebibliography{\DeclareRobustCommand{\VAN}[3]{##3}\VANthebibliography}

\usepackage[pdftex]{graphicx}	
\usepackage{amsmath}	

\usepackage[normalem]{ulem} 

\usepackage{hyperref}

\newcommand{\feiil}{[Fe\,{\sc ii}]\,$1.2570\,\mu$m}
\newcommand{\feii}{[Fe\,{\sc ii}]}

\newcommand{\hml}{H$_2\,2.1218\,\mu$m}
\newcommand{\brg}{Br$\gamma$}
\newcommand{\pab}{Pa$\beta$}

\newcommand{\kms}{km\,s$^{-1}$}
\newcommand{\cmm}{cm$^{-3}$}

\title[Gemini NIFS AGN survey - Gas Kinematics]{Gemini NIFS survey of feeding and feedback in nearby Active Galaxies - V. Molecular and Ionised Gas Kinematics}

\author[M. Bianchin et al.]{
M. Bianchin$^{1}$\thanks{E-mail: marina.bianchin@acad.ufsm.br},
R.~A. Riffel$^{1}$,
T. Storchi-Bergmann$^{2}$,
R. Riffel$^{2}$,
D. Ruschel-Dutra$^{3}$,
C.~M. Harrison$^{4}$,
\newauthor
L.~G. Dahmer-Hahn$^{5,6}$,
V. Mainieri$^{7}$,
A.~J. Schonell$^{8}$,
N. Z. Dametto$^{9}$ \\
$^{1}$Universidade Federal de Santa Maria, Departamento de F\'isica, Centro de Ci\^encias Naturais e Exatas, 97105-900,
Santa Maria, RS, Brazil\\
$^{2}$Instituto de F\'isica, Universidade Federal do Rio Grande do Sul, Av. Bento Gon\c calves 9500, 91501-970, 
Porto Alegre, RS, Brazil \\
$^{3}$Departamento de F\'isica, Universidade Federal de Santa Catarina, P.O. Box 476, 88040-900, Florian\'opolis, SC, Brazil\\
$^{4}$School of Mathematics, Statistics and Physics, Newcastle University, NE1 7RU, UK\\
$^{5}$Shanghai Astronomical Observatory, Chinese Academy of Sciences, 80 Nandan Road, Shanghai 200030, China\\
$^{6}$Laborat\'orio Nacional de Astrof\'isica. Rua dos Estados Unidos, 154, CEP 37504-364 Itajub\'a, MG, Brazil \\
$^{7}$European Southern Observatory, Karl-Schwarzschild-Strasse 2, Garching bei M\"unchen, Germany\\
$^{8}$ Instituto Federal de Educa\c c\~ao, Ci\^encia e Tecnologia Farroupilha, BR287, km 360, Estrada do Chapad\~ao, 97760-000, Jaguari - RS, Brazil\\
$^{9}$Centro de Astronom\'ia (CITEVA), Universidad de Antofagasta, Avenida Angamos 601, Antofagasta, Chile\\
}

\date{Accepted 2021 November 26. Received 2021 November 25. in original form 2021 August 21}

\pubyear{2021}
\usepackage{newtxtext,newtxmath}
\begin{document}
\label{firstpage}
\pagerange{\pageref{firstpage}--\pageref{lastpage}}
\maketitle

\begin{abstract}
We study the gas distribution and kinematics of the inner kpc of six  moderately luminous ($43.43\leq\log{L_{\rm bol}}\leq44.83$) nearby ($0.004\leq z\leq0.014$) Seyfert galaxies observed with the Near-infrared Integral Field Spectrograph (NIFS) in the J($1.25\mu$m) and K($2.2\mu$m) bands. We analyse  the most intense emission lines detected on these spectral wavebands: \feiil\ and \pab, which trace the ionised gas in the partially and fully ionised regions, and \hml, that traces the hot ($\sim 2000$\,K) molecular gas.  The dominant kinematic component is rotation in the disc of the galaxies, except for the ionised gas in NGC\,5899 which shows only weak signatures of a disc component. We find ionised gas outflow in four galaxies, while signatures of H$_2$ outflows are seen in three galaxies. The ionised gas outflows display velocities of a few hundred \kms, and their mass-outflow rates are in the range $0.005-12.49$\,M$_{\odot}$yr$^{-1}$. Their kinetic powers correspond to $0.005-0.7$ per cent of the AGN bolometric luminosities. 
Besides rotation and outflows signatures in some cases, the H$_2$ kinematics reveals also inflows in three galaxies. The inflow velocities are $50-80$\,\kms and the mass inflow rates are in the range $1-9\times10^{-4}$\,M$_{\odot}$yr$^{-1}$ for hot molecular gas. These inflows might be only the hot skin of the total inflowing gas, which is expected to be dominated by colder gas. The mass inflow rates are lower than the current accretion rates to the AGN, and the ionised outflows are apparently disturbing 
the gas in the inner kpc.

\end{abstract}




\section{Introduction}
\label{sec:introduction}

The presence of a supermassive black hole (SMBH) at the centre of every galaxy with a spheroidal component is a well-accepted paradigm of extragalactic astronomy \citep{kormendy13, bosch16}{, 
although the absence of a bulge does not imply the abscence of a SMBH \citep{kormendy11}}. Correlations between the SMBH and host galaxy properties, e.g. the $M-\sigma_{\star}$ relation \citep{magorrian98,ferrarese00,gebhardt00,caglar20}, have been studied over more than two decades and such correlations indicate that the growth of SMBHs and their host galaxies are coupled. The galaxy, or interactions with neighbouring galaxies, provides the fuel necessary to feed the SMBH \citep{sb19}
giving origin to active galactic nuclei (AGN). Once triggered, the AGN then release energy in the form of radiation, gas outflows or jets of particles which can affect the evolution of 
the host galaxy creating the so called AGN feedback \citep[e.g.][]{fabian12}.

Galactic scale feedback from AGN is now understood as one of the main mechanisms acting on star formation suppression of a galaxy \citep{harrison17}, transforming it from star-forming to quiescent. Also, the addition of AGN feedback into cosmological models explains the observed luminosity function of galaxies \citep{benson03} and reproduces a realistic galaxy population \citep{bower06,tng50_1,tng50_2}. 

Theoretical models do predict that AGN driven outflows can suppress the star formation in the host galaxy \citep[][]{granato04,zubovas17, costa20} and the relativistic jets can disturb the gas in the inter-stellar medium (ISM) preventing star formation \citep{mukherjee18}. {
On the other hand, depending on the power of the AGN feedback, }some simulations also predict an enhancement of the star formation \citep{hopkins12, nayakshin12, zubovas13, bieri16, zubovas17} or even new stars being formed inside the outflow \citep{ishibashi12,zubovas13,el-badry16}. 
Due to the lack of strong observational constraints, the AGN feedback is usually included in the simulations in an {\em ad-hoc} way \citep{tng50_1}. 

The recent review by \citet{veilleux20} states that molecular outflows ($T<10^4$\,K) are the main drivers of negative feedback, i.e. quenching of star formation in a galaxy. A way to search for molecular gas outflows is by observing the H$_2$ ro-vibrational transitions in the Near-IR using integral field spectroscopy \citep{emonts17,rogemar_N1275,rogemar_cygnus}, where the hot ($\sim2000$\,K) molecular phase of the gas can be mapped even though it represents just the ``tip of the iceberg'' (only the heated part) of a much larger cold gas reservoir, which is usually mapped and studied through CO emission lines detected at millimetric wavelengths. Although evidence has been found of cold molecular gas outflows that could, in principle, suppress star formation \citep{lutz20}, the estimate of the mass outflow rate relies on the conversion factor from CO to H$_2$ which is highly uncertain \citep{veilleux20}. Thus, it is important to directly map the H$_2$ emission, even if this is only possible in its hottest phase, probed by the near-IR emission lines. 

Another key aspect of AGNs is how the SMBH becomes active and remains at this phase through time. \citet{sb19} revised the main processes involved in the SMBH feeding. 
At the largest scales, galaxy mergers \citep[e.g.][]{hopkins14,sb18} and chaotic cold accretion \citep{gaspari13,suzi17}, also called cooling flows, are important phenomena. In the case of galactic scales, {
disc instabilities cause the gas to lose angular momentum} and move towards the centre. The bars can also trap the gas in rings surrounding the galaxy centre, but they do not seem to feed the AGN directly, as no difference between the accretion rates of barred and non-barred galaxies is observed \citep{galloway15}. {
Using HST imaging, \citet{martini03b} showed that the presence of dust is common among AGN hosts \citep{martini03b} and it can be related to the feeding of the SMBH, but no difference is found comparing active and non-active samples. The time scale for the flow of gas through the dust lanes to centre may be larger than the AGN activity cycle, explaining the lack of significant difference of dust distribution between active and non-active galaxies. In a subsequent work, \citet{simoes-lopes07} have shown that, although there is indeed no difference between AGN and non-AGN for late-type galaxies, there is a marked difference in early-type galaxies: most early type AGN show dust, compared to only 25\% of the control galaxies. This suggests an external origin for the gas in early-type AGN.}

More direct observations of the AGN fuelling are based on studies of the gas dynamics in the inner kiloparsec of AGN hosts,  using integral field spectroscopy which has been successful in mapping gas streaming towards the centre of galaxies \citep[e.g.][]{fathi06,sb07,rogemarN4051,rogemarM79,muller-sanchez09,Diniz15,allan17}. Signatures of gas inflows are observed in multi-gas phases: (i) the cold molecular gas ($\sim100$\,K) can be traced at millimetric wavelengths with facilities like the Atacama Large Millimeter Array (ALMA) using spatially resolved observations of the CO transitions \citep[e.g.][]{combes13}; (ii) the hot molecular gas ($\sim2000$\,K) that is observed in the Near-IR and can be mapped through the H$_2$ emission lines in the K band \citep{rogemarM79,Diniz15}, where the strongest is the H$_2$(1-0) S(1)$2.1218\,\mu$m emission line; (iii) the hot ionised gas ($\sim10,000$\,K) using optical emission lines like the [N{\sc ii}] \citep[e.g.][]{schnorr-muller17}. The inflow rates are typically $\sim1$M$_{\odot}$yr$^{-1}$, which are three orders of magnitudes higher than the SMBH accretion rates, meaning the gas can feed the AGN and accumulate at the centre and form stars over an AGN duty-cycle \citep{sb19}. Other studies of the inner region of active galaxies found that the molecular hydrogen emission usually originates from the gas in the disc \citep{barbosa14, rogemar_n5929}, meaning its kinematics tend to be similar to the stellar one, but in some cases streaming motions towards the centre may also be observed \citep{rogemarN4051,rogemarM79,Diniz15}. 
Hot H$_2$ outflows are also observed in some nearby AGN hosts at scales of a few hundred parsecs \citep[e.g.][]{davies14, ramos-almeida17, rogemar_N1275,rogemar_cygnus}.

Our group (the AGNIFS group) has been using observations with the Gemini Near-Infrared Integral Field Spectrograph (NIFS) to map the gas and stellar properties in the inner kiloparsec of nearby active galaxies. The main sample is composed of 20 objects and is described in \citet{rogemar_sample}. So far, 16 objects have been observed with NIFS or have VLT-SINFONI archival data. {
The observation of one galaxy is on queue for this semester and we expect to conclude the observations of the remaining three objects in 2022, before NIFS is decommissioned from the observatory.}
The gas kinematics of both ionised and molecular gas usually present a rotating disc component. Outflows are commonly observed in ionised gas and scarce in molecular gas which, in some cases, besides the rotation, presents signatures of inflows \citep{sb_N4151_kin, M1066KIN, rogemarM79, rogemar_m1157, rogemar_n5929letter, Schonell14, Diniz15, SchonellN5548, astor19, Diniz19, Gnilka20,rogemar_N1275}.
Our previous work \citep{astor19} shows flux distribution and kinematic maps of the same galaxies we are studying here. The biggest difference between this paper and the previous is that now we disentangle the different emission components (see Sec.\,\ref{sec:measurements}) and quantify the non-rotational motions, like outflows and inflows. Such features were just suggested, and not investigated in deeper detail, in \citet{astor19}.
    

In this paper, we continue our series of works aimed at investigating the AGN feeding and feedback processes. We analyse the ionised and molecular gas kinematics of six active galaxies observed with Gemini NIFS in the J and K bands. In Section~\ref{sec:sample} we present our sample and a summary of the observations and the data reduction procedure. Sec.~\ref{sec:measurements} presents the method used to fit the emission-line profiles, while in Sec.~\ref{sec:results}  we present the resulting flux, velocity and velocity dispersion maps. We compare our results with those from the literature and further analyse our results in Sec.\ref{sec:discussion}, summarising the impact of the gas inflows and outflows on the galaxies in Sec.\,\ref{sec:feed}. Our conclusions are presented in Sec.~\ref{sec:conclusion}.
Throughout this paper we adopt $h=0.7$, $\Omega_m=0.3$ and $\Omega_{\lambda}=0.7$ cosmology.

\section{The Sample and Data}
\label{sec:sample}
\subsection{The sample of active galaxies}
The galaxies in this paper are part of the AGNIFS sample \citep{rogemar_sample},
composed of Seyfert galaxies observed with the Near-infrared Integral Field Spectrograph (NIFS) on the Gemini North Telescope. This sample was selected by adopting the following criteria: (i) X-ray luminosity $L_{\rm X}>10^{41.5}$\,erg\,s$^{-1}$ in the 14-195\,keV band in the 105-month Swift BAT catalogue \citep{BAT105}; (ii) $z\leq0.015$; (iii) $-30^{\circ}<\delta<73^{\circ}$, making them accessible to NIFS. An additional criterion -- the presence of extended [O{\sc iii}] optical emission previously detected in the galaxies (see \citet{rogemar_sample} and references therein) -- was added to guarantee these sources present extended emission from the ionised gas in the Near-IR. The hard X-ray luminosities are particularly suited to the selection of AGNs because, in principle, this radiation is not obscured in the line of sight as would happen in optical wavebands up to Compton thick densities \citep{ricci15}. The redshift criterion guarantees we reach tens of parsec in spatial resolution, necessary for resolving inflows and compact outflows.


In this work, we study six galaxies selected from the AGNIFS sample: NGC\,788, NGC\,3516, NGC\,5506, NGC\,3227, NGC\,5899 and Mrk\,607. These galaxies are among the 50\% most luminous ones without previous studies of their kinematics using NIFS. The other ten galaxies have already been studied by our group  or are currently being analysed (Riffel et al., in preparation).
They cover a luminosity range of $43.4<\log{L_{\rm bol}}(\rm erg\,s^{-1})<44.8$ and a redshift range of $0.004<z<0.014$. 
We obtained J (1.25\,$\mu$m) band data cubes for all galaxies, K (2.20\,$\mu$m) for NGC\,788, NGC\,3516 and NGC\,5506, and K$_{\rm long}$ (2.30\,$\mu$m) for NGC\,3227, NGC\,5899 and Mrk\,607. In the K$_{\rm long}$ grating, some molecular hydrogen transitions are not available, but this does not affect the analysis we perform here, which is focused on the analysis of the strongest emission lines from the molecular and ionised gas. 

 The NIFS data of these galaxies has been previously used in two studies. In \citet{astor19}, we presented the data and gas distribution and kinematics based on emission-line fitting with Gauss-Hermite functions (for details on the differences between the present and previous fitting methods see Sec.\,\ref{sec:measurements}). We also present the spatial distributions of ionised and molecular gas. In \citet{rogemar_excitation} we analyse the gas excitation concluding that thermal processes, such as shocks due to gas outflows, are the main source of excitation to both \feii\ and H$_2$.

\subsection{Near-IR observations and data}

\begin{figure*}
    \centering
    \includegraphics[width=\textwidth]{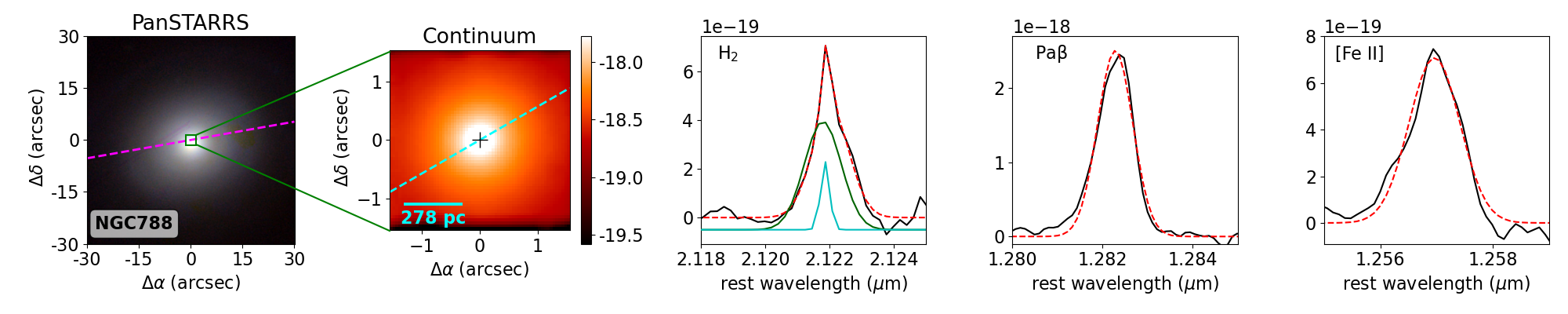}
    \includegraphics[width=0.95\textwidth]{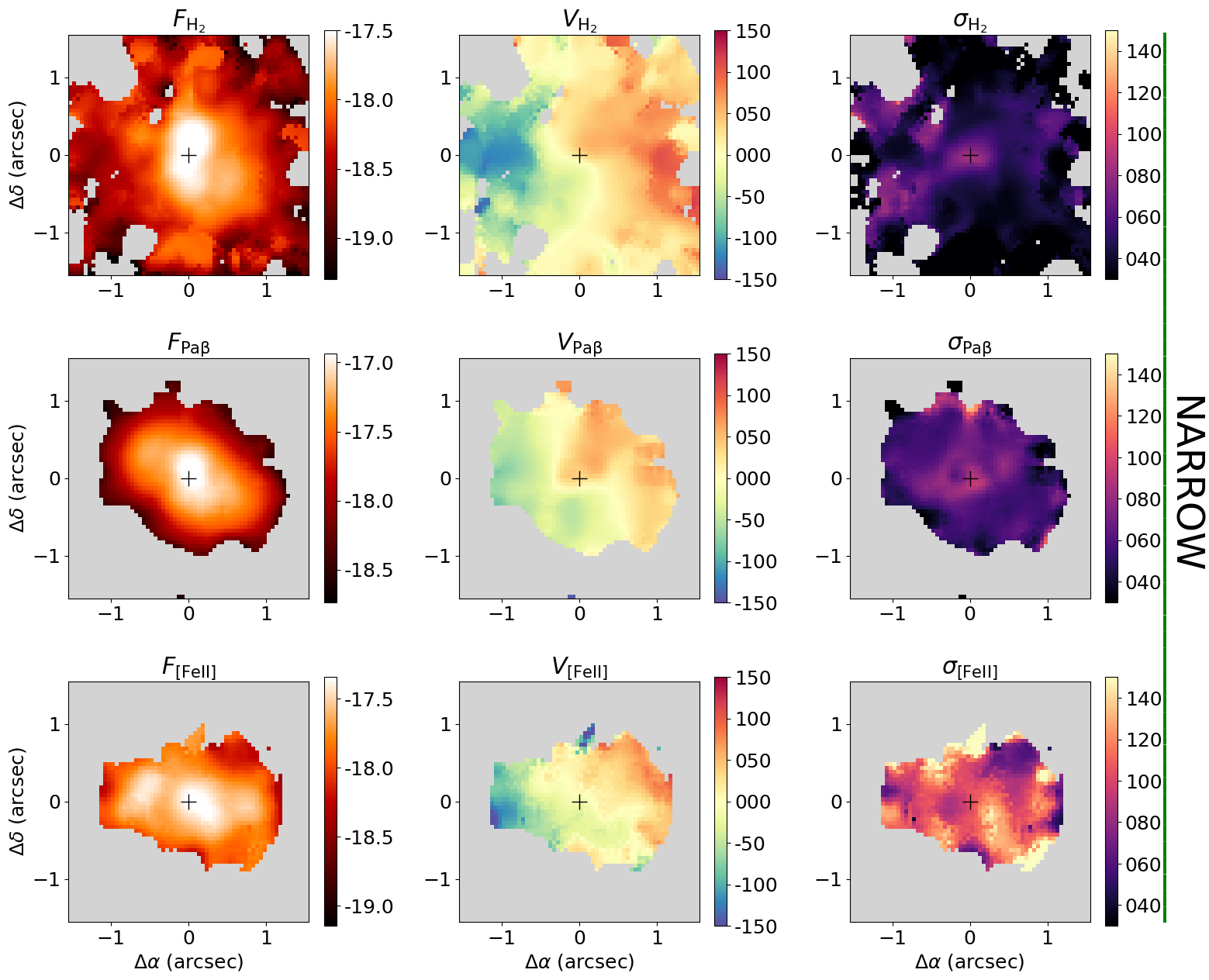}
    \caption{{\bf NGC\,788:} First row shows, in the left panel a PanSTARRS large scale ($1\times1$\,arcmin$^2$) colour image in the y, i and g bands. The green square indicates the NIFS FoV ($3\times3$\,arsec$^2$) which is zoomed in in the continuum image from a  100\AA\ window centred at $2.14\mu$m in the K band shown in the adjacent panel. The magenta and cyan dashed lines indicate the orientation of the galaxy major axis and stellar kinematics position angle \citep{rogemar_stellar}, respectively. Examples of the H$_2$, \pab\ and \feii\ profile fits are shown in the three panels to the right. The continuum contribution is subtracted from the observed (black continuous line) and modelled (red dashed line) spectra. In the H$_2$ panel the narrow component is represented in green and the spurious sky component in blue. Bottom three rows: 2D maps obtained from the Gaussian fit to the emission lines. The first column shows the flux distribution in units of erg\,s$^{-1}$\,cm$^{-2}$\AA$^{-1}$ for the H$_2\,2.1218\mu$m, Pa$\beta$ and \feiil. The second and third columns show the velocity and velocity dispersion maps in \kms 
    The black cross indicates the position of the nucleus 
     (peak of the continuum flux as in \citet{astor19}).
    The grey regions correspond to positions with line amplitudes smaller than 3$\sigma$ of the continuum flux adjacent to the line or with spurious measurements. In all maps the North is up and East to the left.}
    \label{fig:maps_N788}
\end{figure*}

\begin{figure*}
    \centering
    \includegraphics[width=\textwidth]{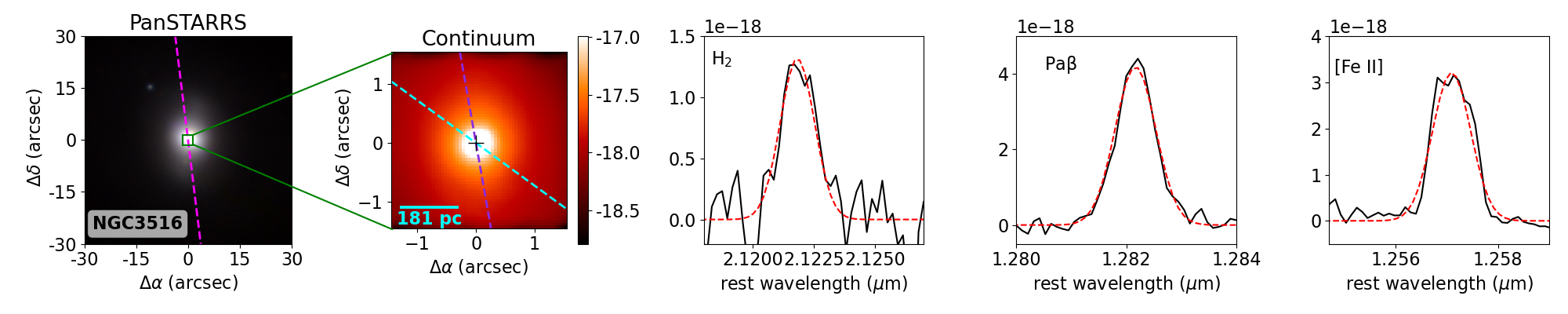}
    \includegraphics[width=0.95\textwidth]{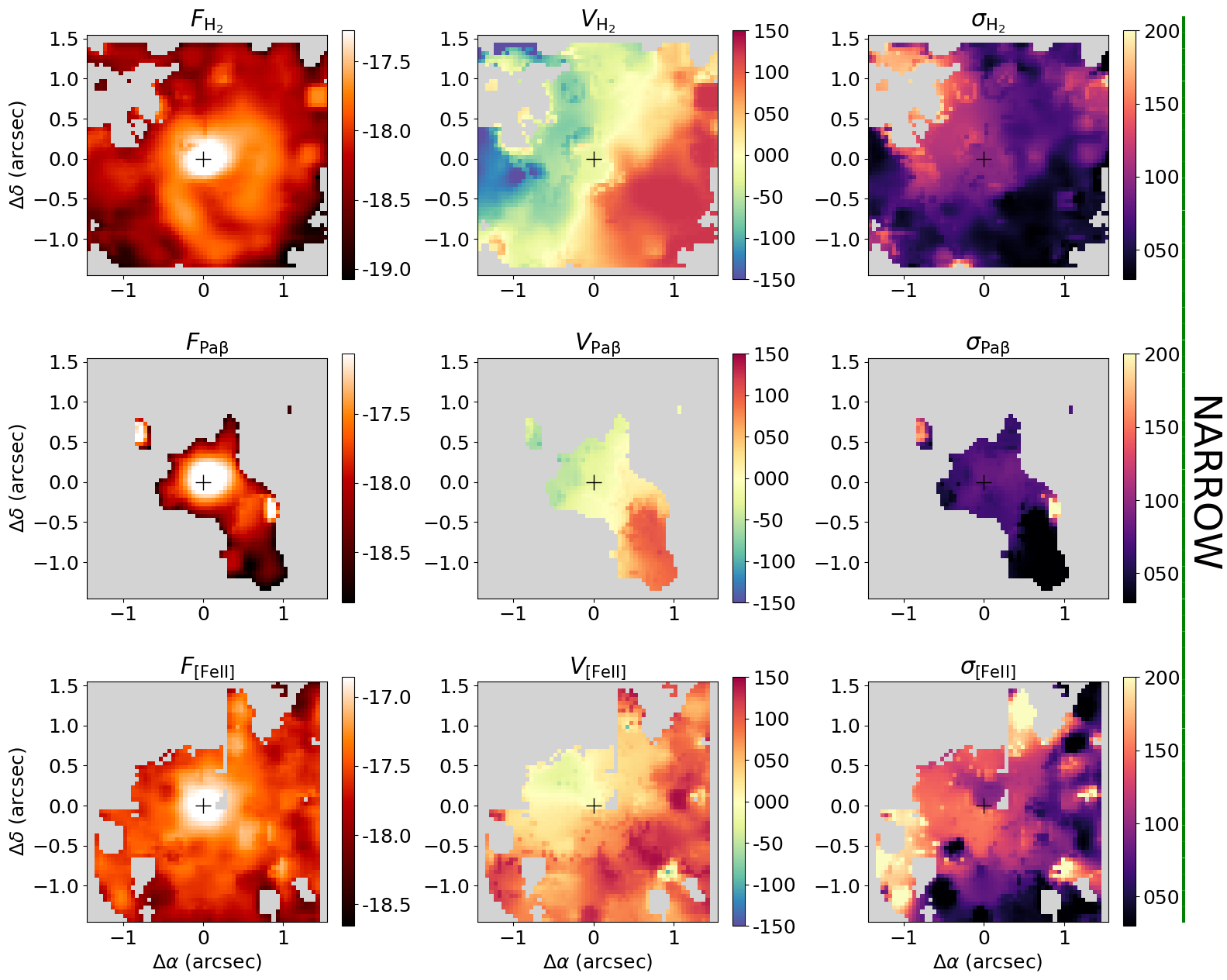}
    \caption{Same as in Fig.\,\ref{fig:maps_N788} for NGC\,3516. The purple line in the continuum map indicates the ionisation axis derived from the [N{\sc ii}] flux distribution \citep{ruschel-dutra21}. The broad line region component to the \pab\ has been subtracted from the observed and modelled spectra.}
    \label{fig:maps_N3516}
\end{figure*}

\begin{figure*}
    \centering
    \includegraphics[width=\textwidth]{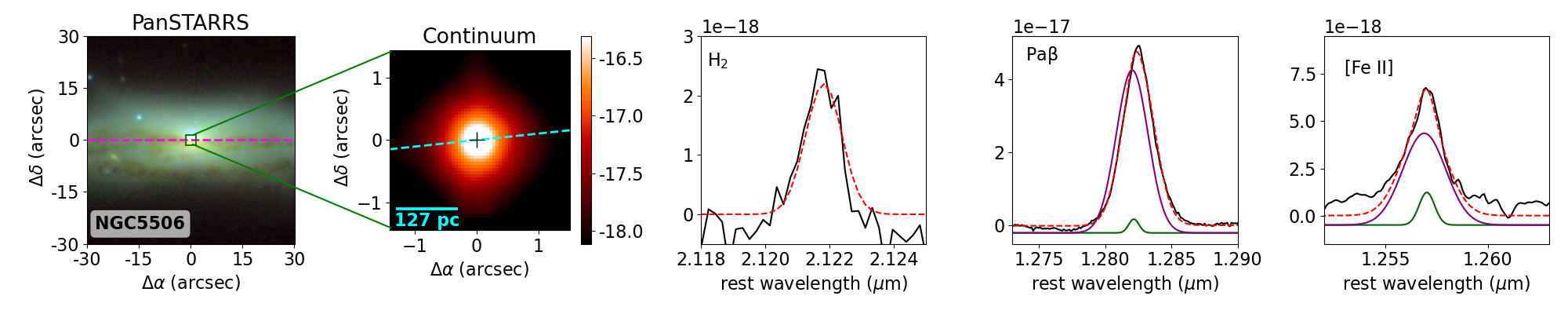}
    \includegraphics[width=0.9\textwidth]{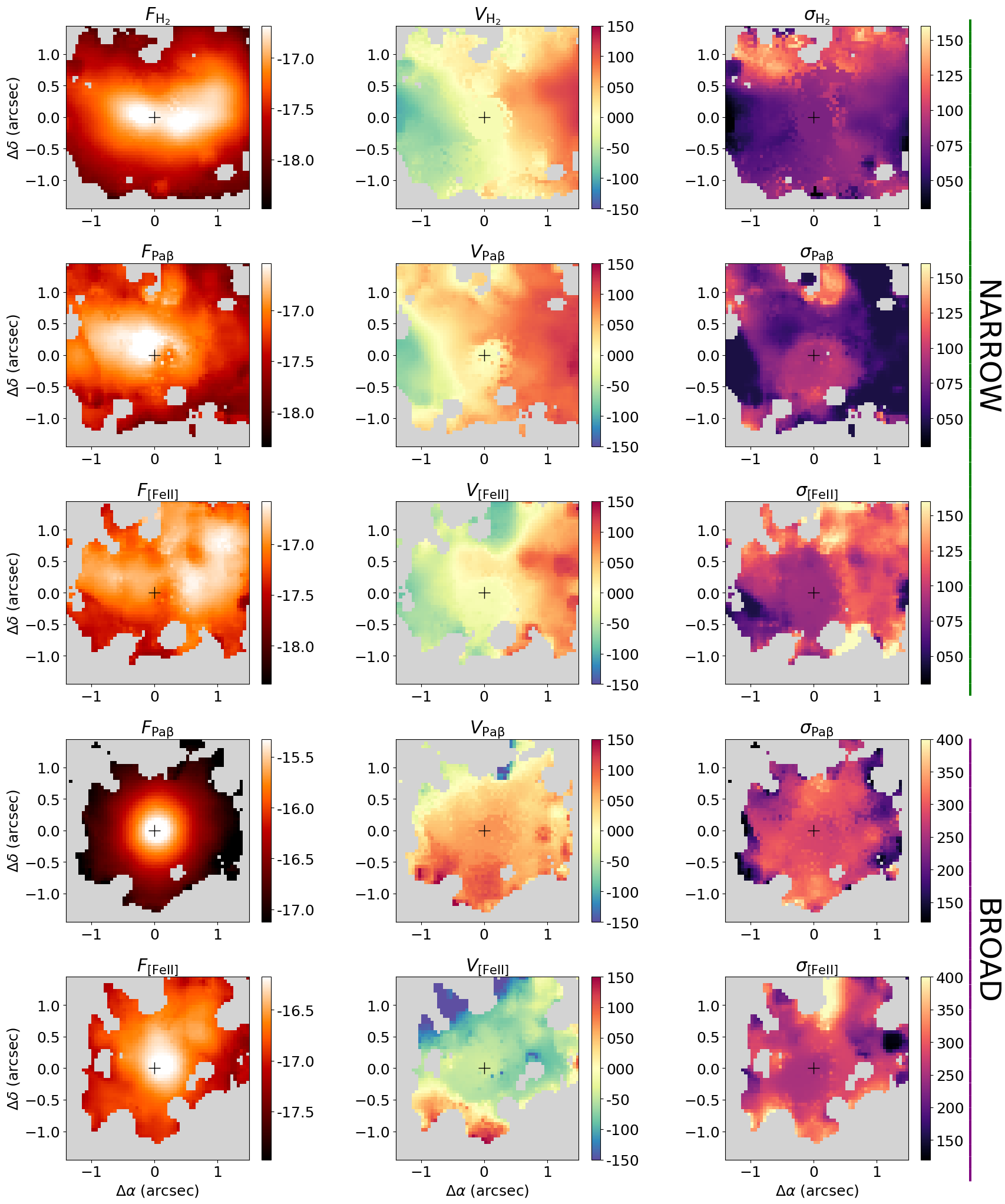}
    \caption{Same as in Fig.\,\ref{fig:maps_N788} for the narrow and broad components of NGC\,5506. In the fits examples the narrow and broad components of \pab\ and \feii\ are represented in green and purple, respectively. The broad line region component to the \pab\ has been subtracted from the observed and modelled spectra.}
    \label{fig:maps_N5506}
\end{figure*}

\begin{figure*}
    \centering
    \includegraphics[width=\textwidth]{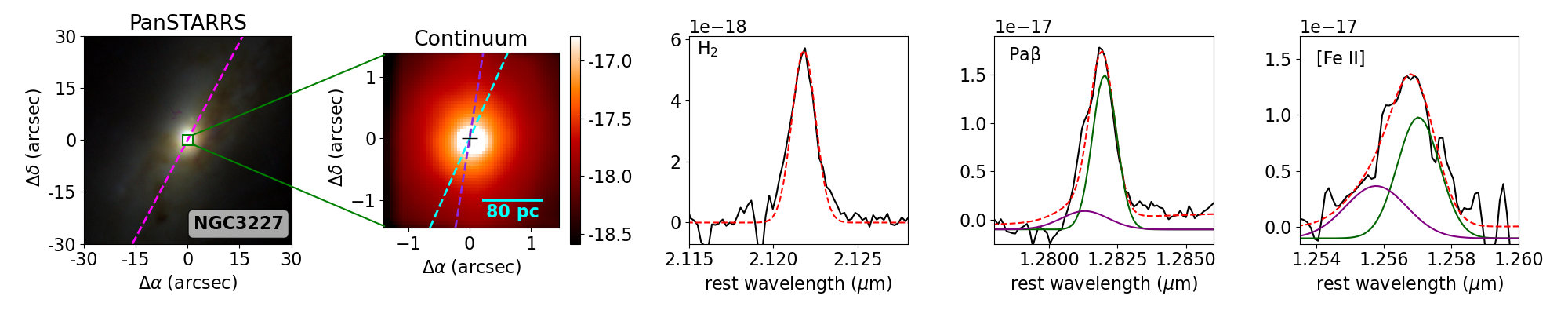}
    \includegraphics[width=0.9\textwidth]{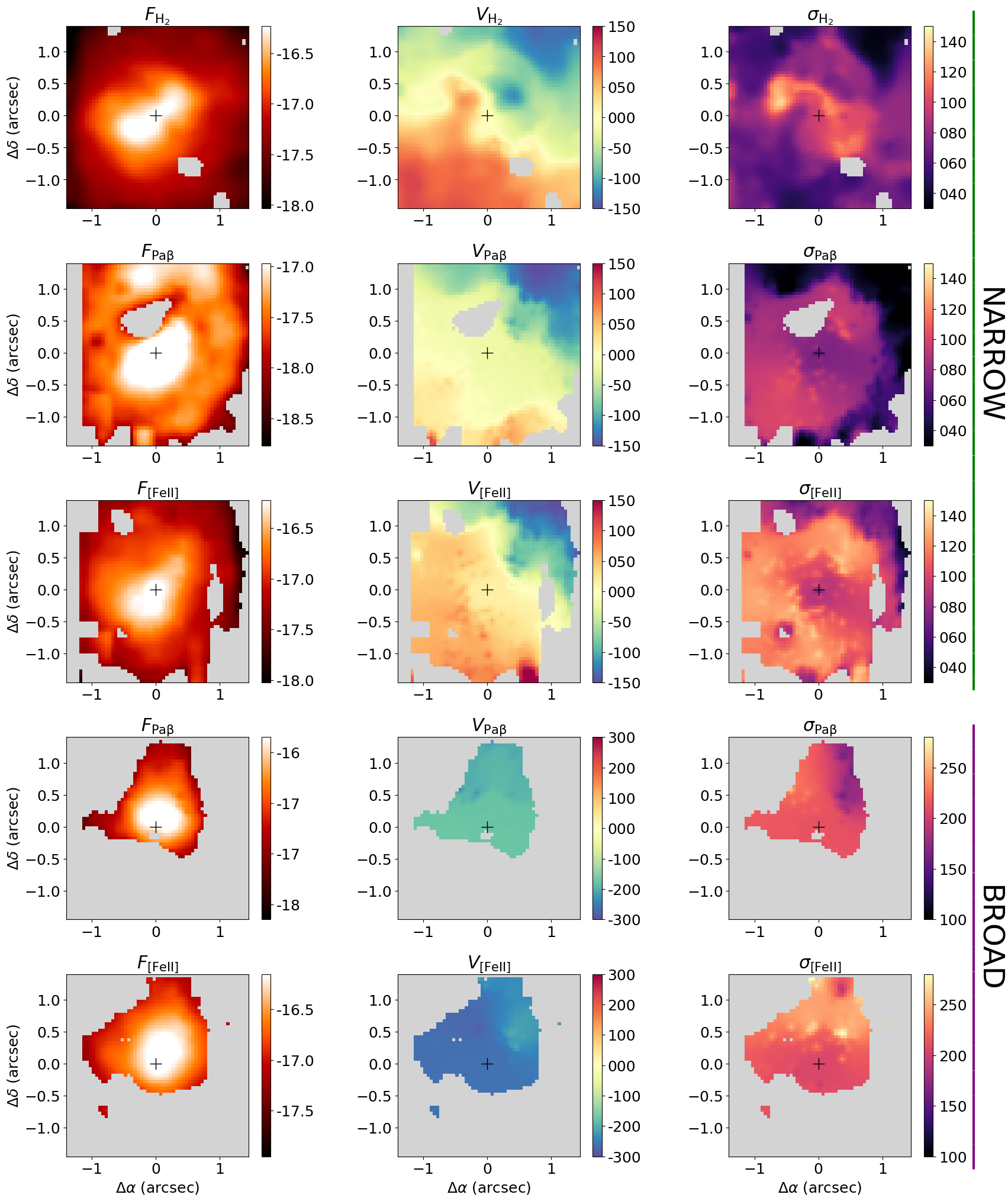}
    \caption{Same as in Fig.\,\ref{fig:maps_N788} for the narrow and broad components of NGC\,3227. In the fits examples the narrow and broad components of \pab\ and \feii\ are represented in green and purple, respectively. The broad line region component to the \pab\ has been subtracted from the observed and modelled spectra.}
    \label{fig:maps_N3227}
\end{figure*}

\begin{figure*}
    \centering
    \includegraphics[width=\textwidth]{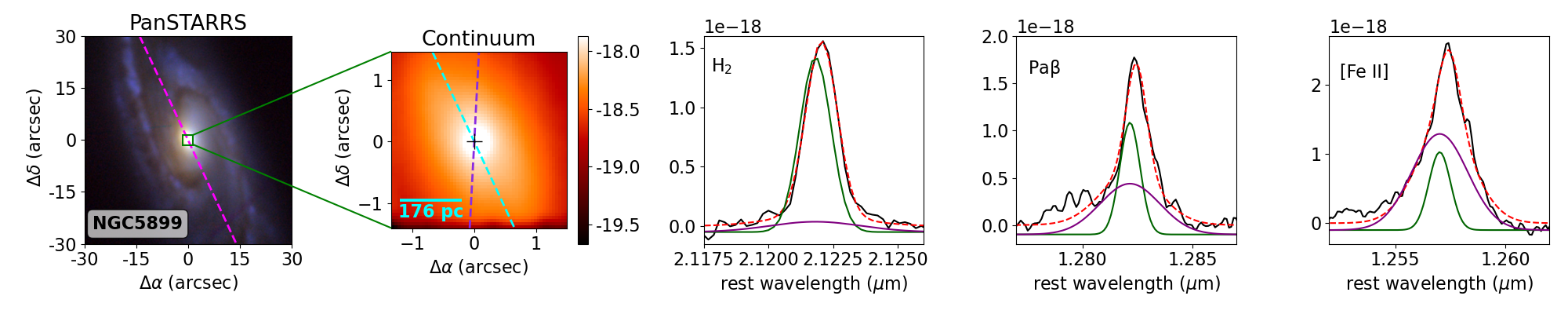}
    \includegraphics[width=0.75\textwidth]{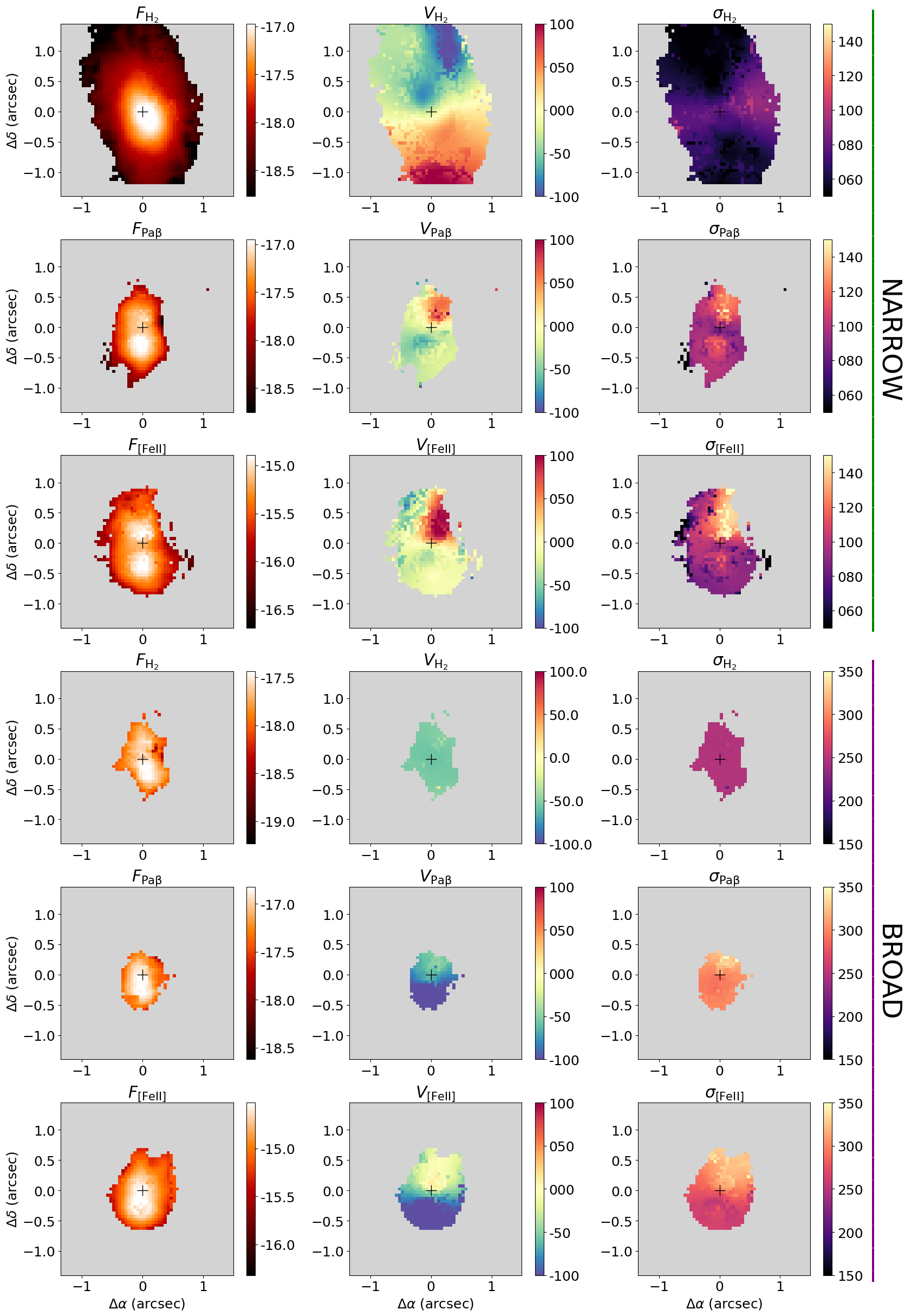}
    \caption{Same as in Fig.\,\ref{fig:maps_N788} for the narrow and broad components of NGC\,5899. In the fits examples the narrow and broad components of H$_2$, \pab\ and \feii\ are represented in green and purple, respectively.}
    \label{fig:maps_N5899}
\end{figure*}

\begin{figure*}
    \centering
    \includegraphics[width=\textwidth]{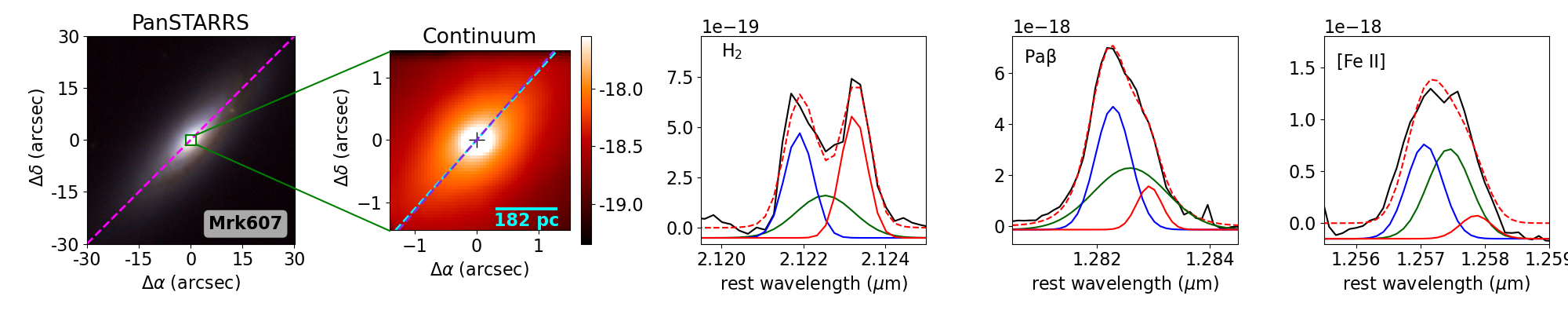}
    \includegraphics[width=0.8\textwidth]{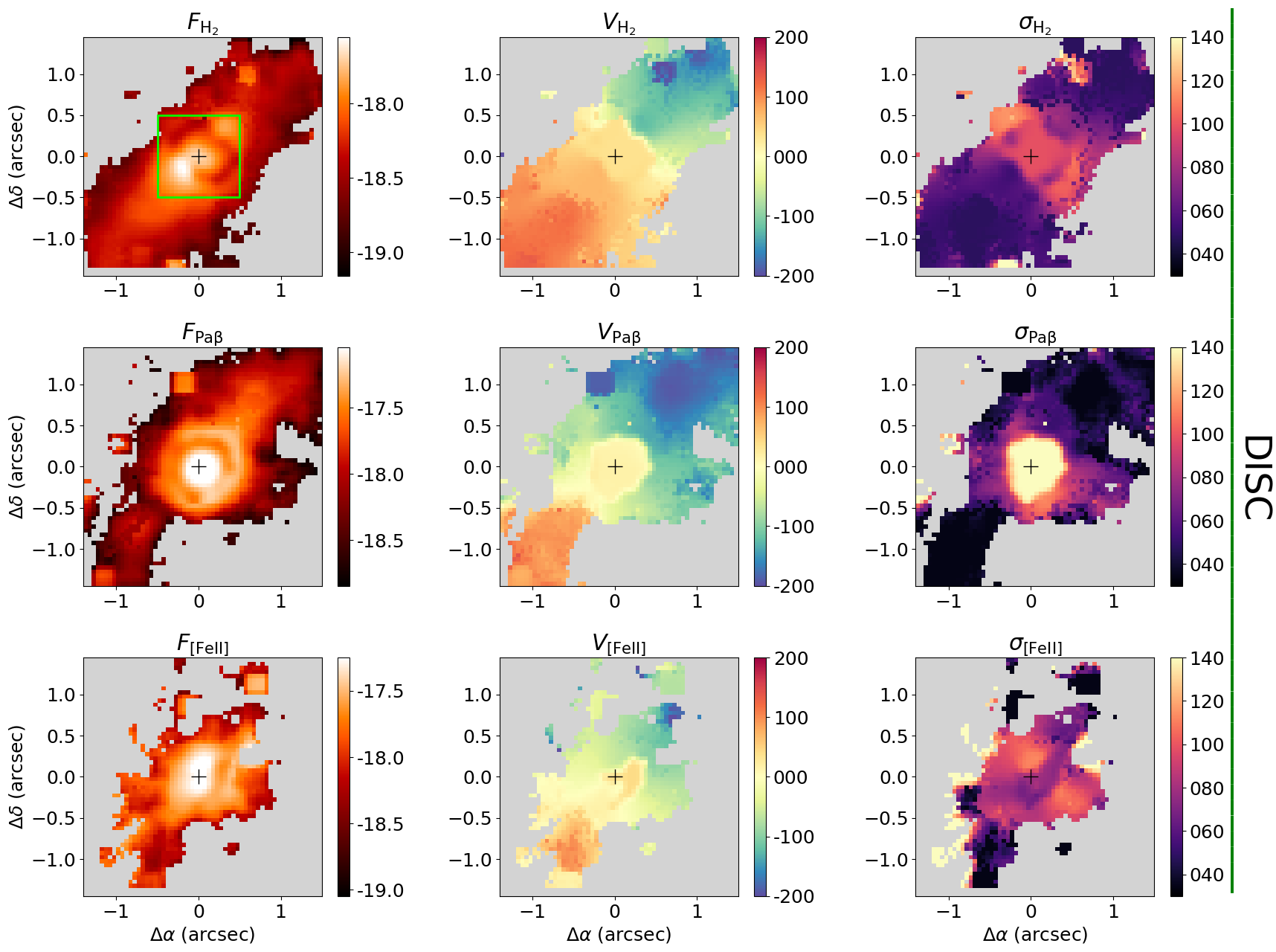}
    \vspace{0.3cm}
    \includegraphics[width=0.489\textwidth]{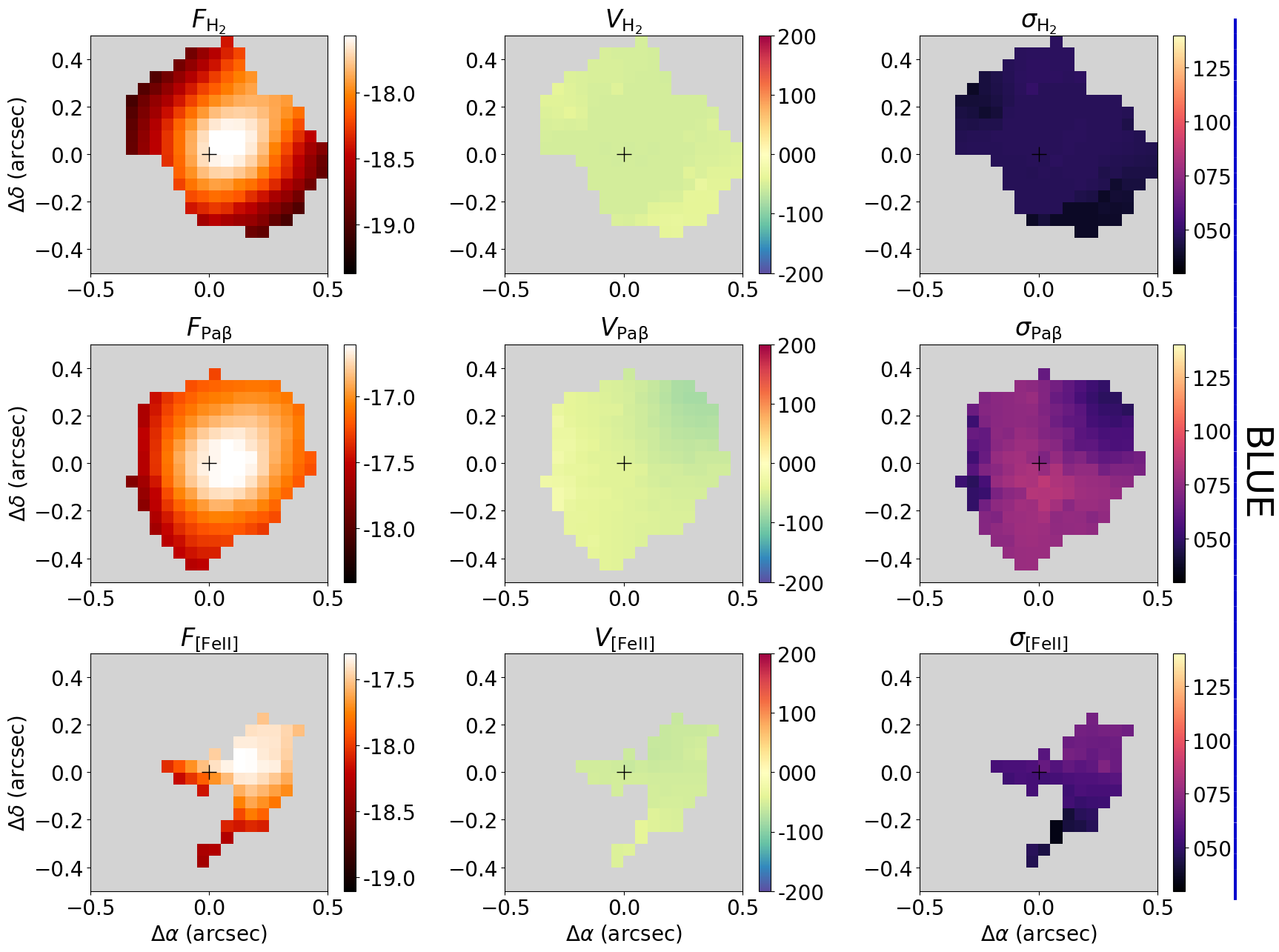}
    \hspace{0.2cm}
    \includegraphics[width=0.489\textwidth]{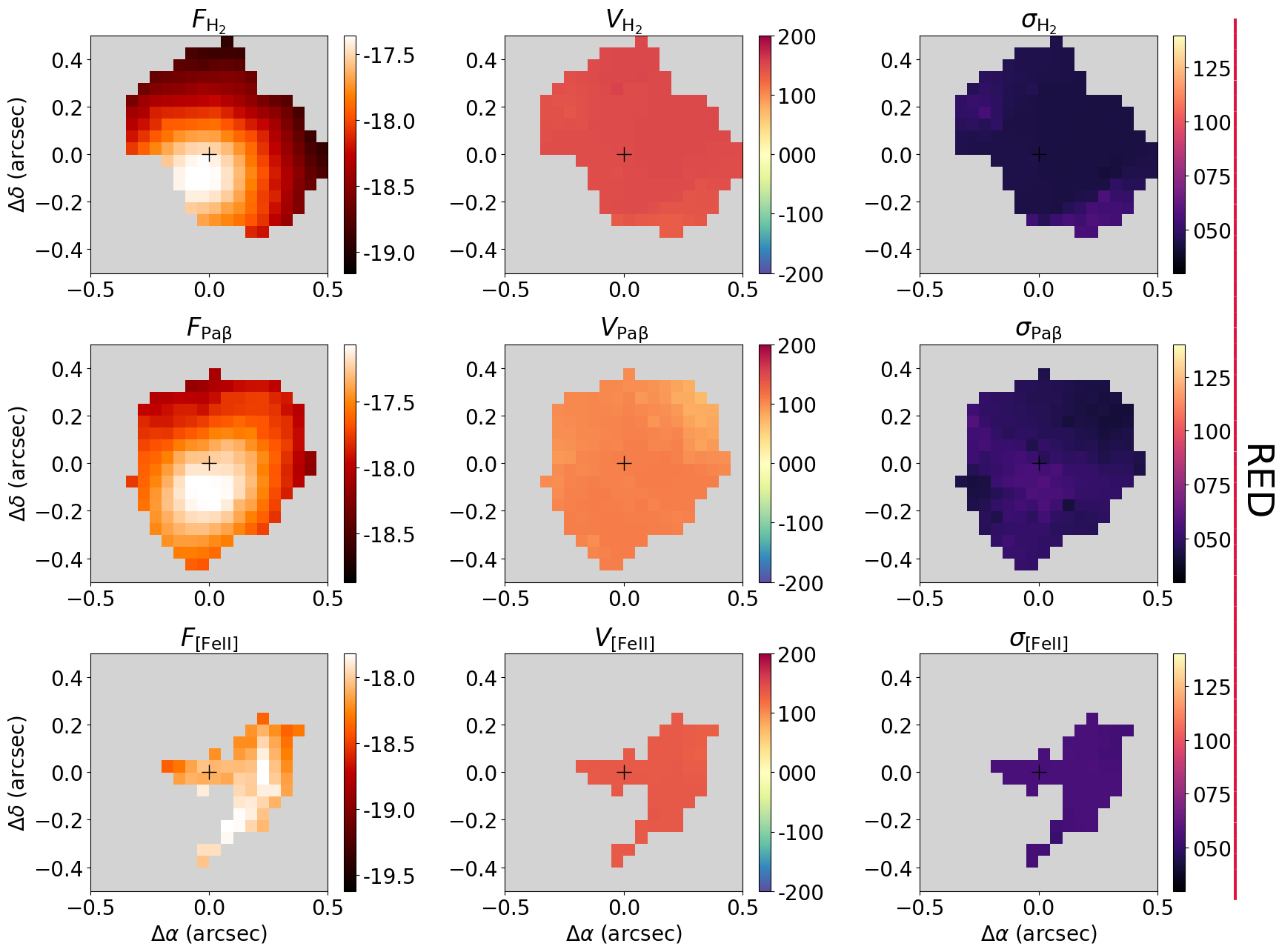}
    \caption{Same as in Fig.\,\ref{fig:maps_N788} for the three profile components in Mrk\,607. In the fits, examples the narrow,  blueshifted and redshifted components of H$_2$, \pab\ and \feii\ are represented in green, blue and red, respectively. 
    The panels 
    showing the emission from the galaxy disc are presented in the top. The emission of the other two Gaussian components, blue and red, are shown in the bottom panels. These components are only relevant at the central region of the maps, which are thus zoomed in to show only the inner $1\times1$\,arcmin$^2$, indicated as a square in the top left panel of the 
    disc component.}
   \label{fig:maps_M607}
\end{figure*}

The NIFS is composed of 29 slitlets with a width of $0.103$\,arcsec and a height of 3.0\,arcsec that together create a field-of-view (FoV) of $3\times3$\,arcsec$^2$. The spatial sampling along each slitlet is $0.042$\,arcsec. At the redshifts of the galaxies studied here, the NIFS FoV covers less than one kiloparsec (see the 1 arcsec scale on the bottom left corner of the continuum maps in Figs.\,\ref{fig:maps_N788}-\ref{fig:maps_M607}). All observations were carried out between 2012 and 2016 (Observing IDs: GN-2012B-Q-45, GN-2013A-Q-48, GN-2015A-Q-3, GN-2015B-Q-29 and GN-2016A-Q-6)  and used the NIFS adaptive optics system ALTAIR (ALTtitude conjugate Adaptive optics for the InfraRed). ALTAIR uses a laser or natural star to guide the observations and correct for the seeing effect.  For the J and K band observations we obtain angular resolutions of $0.13-0.15$\,arcsec and $0.12-0.18$\,arcsec, respectively, that correspond to a scale of $19-36\,$pc and $22-36\,$pc in the J and K bands at the distance of the galaxies. The angular resolutions were estimated from the FWHM of the telluric standard star for the type 2 AGN and from the flux distributions of the broad components of the Pa$\beta$ and Br$\gamma$ emission lines for the type 1 objects.  The spectral resolutions of $\sim1.8$\,\AA, in the J band, and $\sim3.2$\,\AA, in the K band, corresponding to $\approx45$\,\kms, make the NIFS data appropriate to 
characterise the gas kinematics.


The data reduction procedure, explained in detail in \citet{astor19}, uses the Gemini NIFS {\sc iraf} packages. Several exposures of each galaxy are performed (see Table\,1 in \citet{astor19}) and, in each of them, 
the main reduction steps are the flat fielding, sky subtraction, s-distortion correction, wavelength and flux calibrations.  
The datacubes for individual exposures were mean combined using the peak of the continuum emission as the reference for astrometry creating a final data cube with a pixel size of 0.05\,arcsec for each band for all galaxies.    

\section{Measurements and two-dimensional maps}

\subsection{Measurements}
\label{sec:measurements}

The J and K bands spectra of nearby active galaxies are rich in emission lines from ionised species and molecular hydrogen \citep{rogerio06}. 
Here we focus our analysis on the strongest emission lines in active galaxies -- \feiil, \pab\,$\lambda 1.282\mu$m and \hml, which represent distinct gas phases \citep{ardila05,rogerio06,rogerio13,rogerio19}. In the specific case of the \hml, this line is also selected because it is less affected by telluric absorptions and is within the observed spectral K-band range in the six galaxies. 

These emission lines have distinct physical origins: the \feiil\ emission originates in the partially ionised region, where the flux of ionising radiation is not sufficient to fully ionise the gas, has an ionisation potential of 7.9\,eV and is a good tracer of AGN outflows \citep{sb_N4151_kin,barbosa14}; the \pab\ is an recombination line, {
that traces the ionised gas emission;
and the H$_2$ emission lines map the hot skin of the densest regions associated with the cold molecular gas.
}
Due to the common origin, \pab\ and Br$\gamma$ emission lines have the same spatial distribution, i.e. trace the same gas. We choose the former because it is more intense than the latter. 



We fitted the emission line profiles in each spaxel using the Python-based package IFSCube that is designed to handle data cubes from IFS data or single spectrum observations \citep{ifscube}. 
To fit the emission lines we used the {\sc cubefit} module which allows the fit of one or more Gaussian or Gauss-Hermite functions to the observed emission line profiles.  Initial guesses and/or bounds for each of the fitting parameters can be provided by the user: the amplitude ($A$), centroid velocity ($V$), velocity dispersion ($\sigma$), and the $h3$ and $h4$ moments specifically for the Gauss-Hermite fit. 

We also removed the continuum contribution with a low order polynomial. As we selected only the region close to the line to be fitted, in almost all of our fits we used a first-degree polynomial to represent the continuum. The exceptions are the third order polynomial in both the J and K band fits of Mrk\,607 and in the K band fit of NGC\,5899, and the fourth-order polynomial for the J band in NGC\,3516.

In our sample, three galaxies are classified as Seyfert 1 \citep{BAT105}: NGC\,3227, NGC3516 and NGC\,5506, meaning the broad line region (BLR) of the AGN is detected. The BLR is identified as a broad component ($\sim 1000$\,\kms) superimposed to the other components of the \pab\ and \brg, in the J and K bands spectral range. {
Since the BLR is not spatially resolved, }their line profiles were fitted by a Gaussian function with fixed velocity and velocity dispersion in all spaxels they appeared, as measured from the nuclear spectrum of each galaxy.  

In previous works with these galaxies different strategies to fit the emission-lines have been adopted. In \citet{astor19} Gauss-Hermite functions were chosen because the emission-line profiles deviated from a Gaussian function. A different approach was adopted in \citet{rogemar_excitation} where up to three Gaussian functions were fitted to each emission-line. Both approaches aimed at the best representation of the observed line profiles and no physical meaning to the components was attributed. 
{
In this paper, we fit up to three Gaussian functions to the profile of each emission line. The criterion to decide which model provides the best representation of the observed spectral features is based on the following procedure. We calculate the mean of the residuals, observed spectra $-$ model, divided by the standard deviation for each emission line within a window of 1000\kms. We then select as the best model the one that gives the lowest value for this residual. 
An special case is Mrk\,607, where the emission line profiles have a clear double peak structure and it has a well-known equatorial outflow \citep{Freitas18}, so our procedure tried to address this.}
Using this strategy and previous results from the literature on these galaxies, we fitted the following number of Gaussian functions to the \feii, \pab\ and \hml\ emission lines:
\begin{itemize}
    \item for NGC\,788 and NGC\,3516, a single Gaussian component can reproduce the observed profiles at all locations. For NGC\,788 a sky feature close to the \hml\ significantly affected the maps presented in \citet{astor19}. {
    We have now taken care of this feature as follows. The line profile was fitted by two components: one representing the H$_2$ emission and the other, with fixed width ($\sigma=20$\,\kms) and centred at 2.1218\,$\mu$m in all spaxels, for the sky contamination. Since the sky feature is just a contamination, we consider that the H$_2$ is reproduced by a single Gaussian function.}
    \item for NGC\,5506 and NGC\,3227, the \feiil\ and \pab\ line profiles were fitted with two Gaussian curves each, while the \hml\ is well reproduced by a single Gaussian function at all locations;
    \item two Gaussian components are needed to reproduce each emission line of NGC\,5899;
    \item  for Mrk\,607, the emission lines are well reproduced at most locations by a single Gaussian curve tracing the emission of gas in the galaxy disc, except for the inner 0.5\,\arcsec where three components are present, one due to the disc and two produced by an equatorial outflow previously observed in spectra of this galaxy \citep{Freitas18}. The same strategy has already been adopted to describe the equatorial outflow in NGC\,5929 \citep{rogemar_n5929}.
\end{itemize}
In the cases where two Gaussians are necessary, one is broad ($\sigma\sim400$\,\kms) and one is narrow ($\sigma\sim100$\,\kms). Examples of the fits, extracted from the nuclear spaxel, are presented in the first line of Figs.\,\ref{fig:maps_N788}-\ref{fig:maps_M607}.

\subsection{Flux and kinemactic maps}
\label{sec:results}

In the top left panel of Figures~\ref{fig:maps_N788}-\ref{fig:maps_M607} we present the PanSTARRS (Panoramic Survey Telescope and Rapid Response System) 
{\it g}, {\it i} and {\it y} composite image \citep{chambers19, flewelling20} in large scale ($1\times1$\,arcmin$^2$) for each galaxy in our sample.  
To the right, we present an image of the continuum, obtained directly from the NIFS data cube in a region free of emission lines, centred at $2.14\mu$m, by calculating the mean of the spectra in a window of 100\,\AA. The dashed lines in these two images indicate the orientation of the galaxy major axis and the orientation of the line of the nodes of the stellar velocity field \citep{rogemar_stellar}.
The other three panels show examples of the \hml, \pab\ and \feiil\ emission line fits, extracted from the spaxel at the peak of the continuum emission. The black continuous line is the observed and the the red dashed line the modelled spectra. For clarity in the plots, the continuum contribution was subtracted from both. When more than one Gaussian function was fitted, we present the individual components as coloured continuous lines: green for narrow, purple for broad and red and blue for the redshifted and blueshifted components in Mrk\,607. 
For all galaxies, the nuclear spaxel presents the most complex emission line profiles and, as can be observed from these figures, the adopted models can properly reproduce the observed profiles.

The emission line fitting was performed over the whole NIFS FoV producing two-dimensional maps for all the fitted parameters. 
In Figs.~\ref{fig:maps_N788}-\ref{fig:maps_M607} we also present 2D maps for the flux, velocity (after the subtraction of the systemic velocity, defined as the velocity obtained by fitting the H$_2$ velocity field by a rotation disc model -- see Sec.\ref{sec:rotation}), and velocity dispersion (corrected for the instrumental broadening) for each Gaussian component fitted to \feii, \pab\ and H$_2$ emission lines of every galaxy in our sample. The name of the galaxy is indicated at the upper left corner and the vertical bars on the right indicate which component the maps are referring to: green to the narrow and purple to the broad components; in the case of Mrk\,607, that has three components, green corresponds to the narrow/disc, blue and red to the blueshifted and redshifted components, respectively. The grey regions in these maps represent locations where the amplitude of the corresponding Gaussian component is lower than 3 times the standard deviation of the continuum level next to each emission line.

In Appendix\,\ref{app:channel} (shown as supplementary file) we present the emission-line channel maps for the three emission lines fitted and analysed here. These channel maps support our interpretation of the molecular and ionised gas kinematics in the galaxies.

\subsubsection{
Highlights from the kinematic maps}
\label{sec:summary}
 A detailed discussion of all the maps, including the flux distributions, and a comparison with previous results for each galaxy is presented in the Appendix~\ref{app:notes_gals}. 
 
 Here we summarise the main features observed in the gas kinematics: (i) the narrow/single component is dominated by rotation, i. e. from gas orbiting in the disc, except the case  of NGC\,5899, for which it is due to ionised outflows; (ii) the broad component in  the ionised gas is detected in three galaxies -- NGC\,5506, NGC\,3227 and NGC\,5899 -- and it is interpreted as tracing gas outflow; (iii) a broad component in the molecular gas is seen only in NGC\,5899 -- interpreted as originating from the interaction between the ionised outflows and the molecular gas in the disc; (iv) an equatorial outflow is detected in Mrk\,607; (v) the narrow component kinematics -- although dominated by rotation -- usually shows deviation, which are mostly present in the ionised gas, but also in H$_2$ in the case of NGC\,3227.

\section{Discussion}
\label{sec:discussion}

\subsection{Rotation models}
\label{sec:rotation}
\label{rotation}
In Figs.\,\ref{fig:maps_N788}-\ref{fig:maps_M607} we observe that the narrow or single component of both molecular and ionised gas velocity fields shows a rotation pattern, but some deviations from pure rotation are also observed. The molecular gas is usually more restricted to the plane of the disc \citep[e.g.][]{reunanen02, rogemar_n5929}, therefore it is prone to having a velocity field dominated by rotational motions. We then fit the narrow/single component 
velocity fields by a rotation disc model, to investigate the residual velocity maps (observed $-$ model) and identify non-circular motions.
The H$_2$ velocity fields are modelled by the following equation, 

\begin{equation*}
\begin{aligned}
    V(R,\psi)=V_{\rm sys}+
    \frac{AR\cos(\psi-\psi_0)\sin\theta\cos^{p}\theta}{\{R^2[\sin^2{(\psi-\psi_0)}+\cos^2{(\psi-\psi_0)}]+C_0\cos^2{\theta}\}^{p/2}},
    \end{aligned}
\end{equation*}
where $R$ is the distance of each spaxel to the nucleus, $V_{\rm sys}$ is the systemic velocity of the galaxy, $A$ is the velocity amplitude, $\psi$ is the position angle of each spaxel in the plane of the sky, $\psi_0$ is the orientation of the line of the nodes, $\theta$ is the inclination of the disc, $C_0$ is a concentration parameter, and the parameter $p$ defines the slope of the rotation curve at the largest radii and is restricted to $1\leq p\leq3/2$.  The equation is based on \cite{bertola91} and has already been used in previous works by our group using similar data \citep[e.g.][]{schnorr-muller14,Diniz15,carine17,carine19}. 

\begin{figure}
    \centering
    \includegraphics[width=0.5\textwidth]{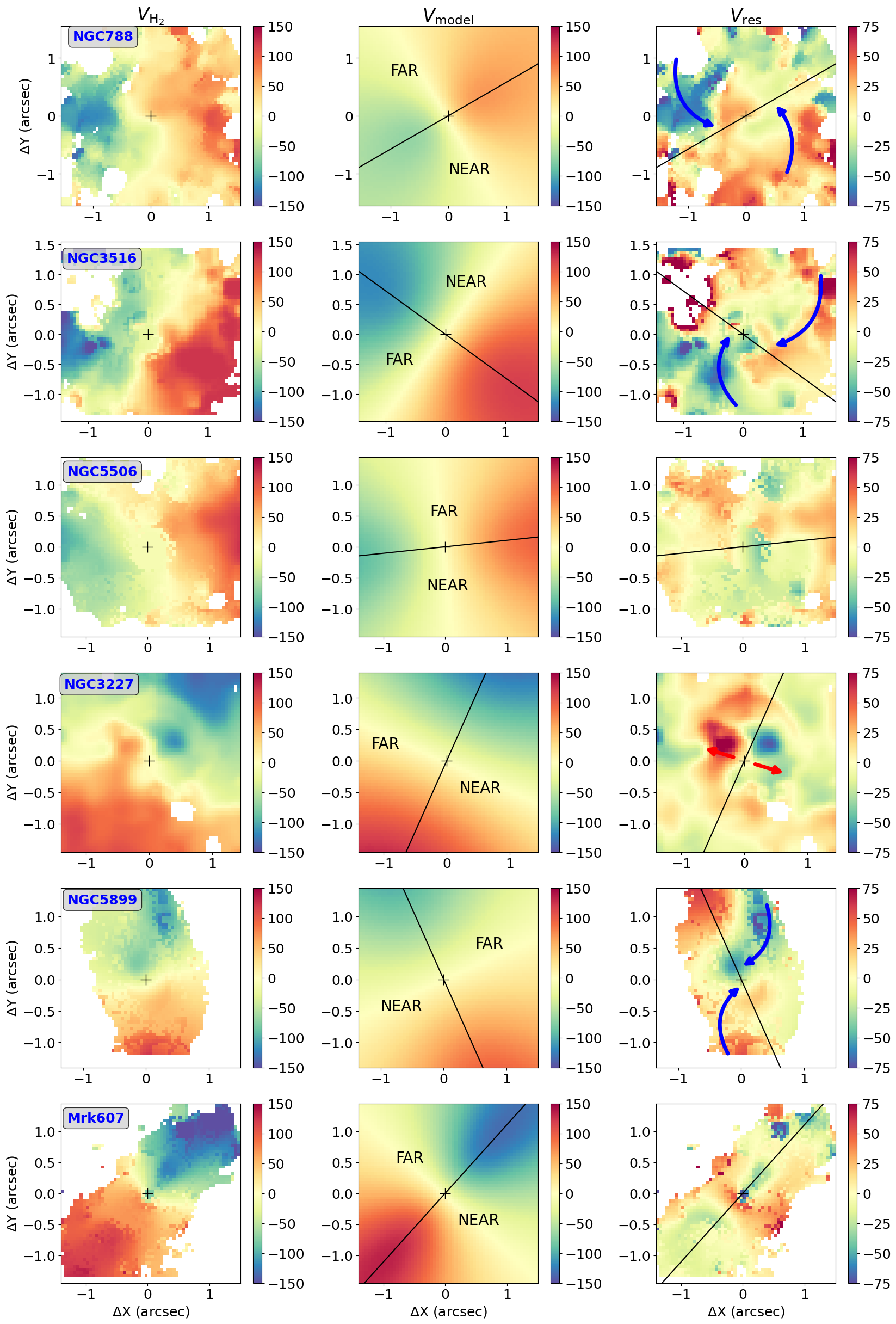}
    \caption{Molecular gas in the disc of the galaxies: The first column shows the H$_2$ velocity field as in Figs.\,\ref{fig:maps_N788}-\ref{fig:maps_M607}. The second column shows the rotation models. The black line indicates the orientation of the line of the nodes ($\psi_0$) as derived by \citet{rogemar_stellar}. The third column shows the residuals ($V_{\rm res}= V_{\rm H_2} - V_{\rm model}$). The black cross marks the position of the AGN. The red arrows indicate residuals interpreted as outflows and blue arrows the inflows.}
    \label{fig:rotation}
\end{figure}

During the fit of the H$_2$ velocity fields
we fix $\psi_0$ and $\theta$ to the values obtained from previous modelling of the stellar velocity fields \citep{rogemar_stellar}. We adopt this procedure because the non-circular motions in the gas kinematics -- e.g. inflows and outflows -- will influence the fit possibly leading to a wrong rotation field, while the stars are a better tracer of circular motions dominated by the gravitational potential of the galaxies.
In the case of Mrk\,607, to take into account the counter-rotation of the gas \citep{rogemar_stellar, Freitas18}, we subtract $180^{\circ}$ from $\psi_0$ and adopt this new value as the orientation of the line of the nodes.
The kinematical centre is assumed as the position corresponding to the peak of the continuum flux in the K band.  We also assume $p=3/2$ 
that our previous studies have shown to reproduce better the inner regions. Previous assumptions of a flat rotation curve ($p=1$) are better suited to the outer regions (beyond the inner kpc probed here) of disc galaxies.  Therefore, $A$, $V_{\rm sys}$ and $C_0$ are the only free parameters of the model. The middle column of Fig.\,\ref{fig:rotation} shows the modelled rotation velocity fields. 

{
We use the colour maps from \citet{martini03a} to determine the far and near sides of the galaxies as indicated in the central panel of Fig.\,\ref{fig:rotation}. The $V-H$ colour map
allows the identification of the most obscured side of the galaxy, which can be identified with its near side. This is due to the fact that the light emitted by the central, most luminous region of the galaxy (bulge and surroundings), due to the galaxy inclination, becomes partially hidden by dust lanes in the near side of the disc, while in the far side this does not happen. 
Thus, the near side is redder and looks dustier than the far side the colour maps of the galaxies. 
In NGC\,3227 the dust lanes are not as clear as in the other galaxies, but our determination agrees with those from previous works \citep{barbosa09, alonso-herrero19}. HST data in $V$ and $H$ band images are not available for NGC\,5899, therefore we use the Pan-STARRS image (Fig.\,\ref{fig:maps_N5899}), assuming the arms are trailing, and from the observed kinematics, determine the southwest as the near side of the galaxy. We use the galaxy major axis determined by \citet{rogemar_stellar} to separate the near from the far sides.}

In order to identify non-circular motions in the H$_2$ kinematics, we build the residual velocity maps ($V_{\rm res}=V_{\rm H_2}-V_{\rm model}$) presented in the third column of Fig.\,\ref{fig:rotation}. If the gas is located in the plane of the disc, blueshifts in the far side and redshifts in the near side indicate that the gas is moving towards the centre of the galaxy. If the opposite is observed, redshift in the far side and blueshifts in the near side, the residuals may be tracing outflows. In all cases, the residual velocities are much smaller than the observed velocity amplitudes, indicating that the velocity fields for the narrow component are dominated by rotation. However, systematic deviations from pure rotation are seen in all galaxies suggesting that besides rotation, the gas displays also non-circular motions. Thus, we summarise and interpret the observed velocity residuals for each galaxy as follows:
\\

\noindent {\em \large NGC\,788:}~~ Mostly blueshifts in the far side and redshifts in the near side of the galaxy with velocities of $\sim$50 km\,s$^{-1}$. Assuming that the gas is in the disc, as indicated by the low $\sigma$ values ($\sim60$ km\,s$^{-1}$) observed (Fig.~\ref{fig:maps_N788}), the residual velocities are consistent with streaming motions towards the centre. {
On the other hand, another possible explanation, as the residuals are more spread over the FoV is that these disturbances in the velocity field originate from the interaction between the possible [Fe{\sc ii}] bipolar outflow with the molecular gas in the disk. } \\ 
\noindent{\em \large NGC\,3516:}~~  The residual map shows mostly redshifts in the near side of the galaxy and blueshifts in its far side. {
The enhanced H$_2$ flux values outside the nucleus are co-spatial with the highest residuals and with the lowest H$_2$ $\sigma$ values (Fig.~\ref{fig:maps_N3516}). The residual maps, H$_2$ flux distribution and velocity dispersion support the presence of streaming motions towards the centre of this galaxy.}\\
\noindent{\em \large NGC\,5506:}~~ Some redshifts are observed in the far side and blueshifts in the near side of the galaxy, approximately along the AGN ionisation axis at position angle of $22^{\circ}$ \citep{fischer13}. {
This result is consistent with the orientation of the ionised outflow observed in Pa$\beta$ and \feii\ (see Appendix \ref{app:n5506})}. The H$_2$ residuals may be tracing the emission of molecular gas located in the outer layers of the outflow seen in the ionised gas. \\
\noindent{\em \large NGC\,3227:}~~  In the central 0.7\,arcsec, redshifts are observed in the far side and blueshifts in the near side of the galaxy, indicating the presence of a compact outflow with velocity amplitudes $\sim80$\,\kms. \citet{davies14} do not see the molecular outflow directly, instead interpret the H$_2$ velocity field distortions as due to an interaction between the ionised outflow and the gas in the disc. On the other hand, a compact molecular outflow seen in the CO(3-2) along the galaxy minor axis with an extension of 70\,pc has been detected for the cold gas \citep{alonso-herrero19}{
, supporting our interpretation. The outflow structure for this galaxy is probably a combination of the scenarios proposed by \citet{fischer13}, where we see inside the ionised outflow bicone (Fig.\,\ref{fig:maps_N3227}), and by \citet{alonso-herrero19} where the molecular emission is more compact and clumpy. }

\noindent{\em \large NGC\,5899:}~~ This galaxy shows the clearest signatures of gas inflows in our sample. A strip of blueshifted gas ($\sim80$\kms) is observed in the northwest side of the nucleus (the far side of the galaxy). Redshifts with similar velocities are seen on the near side of the disc. Both, blueshifts and redshifts, are seen in regions with low H$_2$ velocity dispersion (Fig.~\ref{fig:maps_N5899}), indicating that the gas is located in the plane of the galaxy and thus consistent with inflows.  \\
\noindent{\em \large Mrk\,607:}~~ We use the single Gaussian fit to obtain the rotation model for this galaxy as it represents the bulk of the emission from the disc. The residuals are small at most locations. The residual map presents an excess of redshifts perpendicular to $\psi_0$, which may be associated with the equatorial outflows observed for this galaxy \citep[][and discussion below]{Freitas18}.

\subsection{Inflow properties}
\label{sec:inflow}
The feeding to the AGN can be measured through kinematic signatures on the velocity maps that indicate the gas is flowing towards the centre. In \hml\ such signatures have been observed in some nearby active galaxies, including a few galaxies of the AGNIFS sample \citep[e.g.][]{rogemarN4051, muller-sanchez09,hicks09,rogemarM79, Diniz15}. {
In Sec.\,\ref{sec:rotation}, we present the H$_2$ residual velocity maps for the six galaxies and all of them present disturbances. For three galaxies, NGC\,788, NGC\,3516 and NGC\,5899, a possible interpretation for the residuals -- excess of blueshifts in the near side and of redshifts in the far side -- is the presence of the gas inflows. Although we cannot rule out some other effect, such as disturbances due to ionised outflows, we adopt here the inflow hypothesis.}


Based on \citet{scoville82} we can estimate the mass of hot molecular gas in the disc component gas from the following equation
\begin{equation}
 \left(\frac{M_{\rm H_2}}{M_\odot}\right)=5.0776\times10^{13}\left(\frac{F_{\rm H_{2}\,2.1218}}{\rm erg\,s^{-1}\,cm^{-2}}\right)\left(\frac{D}{\rm Mpc}\right)^2,
\label{mh2}
\end{equation}
where $F_{\rm H_{2}\,2.1218}$ flux of the narrow component of \hml\ summed over the whole FoV and $D$ is the redshift distance of the galaxy.  This equation is valid for gas in local thermodynamic equilibrium ($T_{\rm vib}\approx2000$\,K) that is consistent with the temperatures we measure for these galaxies \citep{rogemar_excitation}. 

The equation above provides the total mass of hot molecular gas in the disc (rotating gas + inflows) that is extended over a region with a radius $r_{\rm d}\approx1$\,arcsec in these three galaxies. This radius encloses an area smaller than the NIFS FoV and is chosen because usually the borders present poorer quality in the emission line fit. The inflowing structures (see $V_{\rm res}$ in Fig.\,\ref{fig:rotation}) appear in pairs with velocities of opposite values in each side of the galaxy resembling double spiral arms ($n_{\rm arms}=2$) that are also clear in the large scale image of NGC\,5899 (Fig.\,\ref{fig:maps_N5899}) where they are more collimated, each with a radius $r\approx0.3$\,arcsec and each with a velocity of $v_{\rm obs}=80$\,\kms. In NGC\,788 and NGC\,3516 the spiral arms are not seen in their large scale images but are observed in the velocity residual maps ($v_{\rm obs}=50$\,\kms), being less collimated than for NGC\,5899, with a radius $r\approx0.5$\,arcsec. The corresponding physical radius for $r_{\rm d}$ and $r$ are presented in Table\,\ref{tab:inflow}.


To obtain the mass inflow rate we assume the gas is moving towards the centre over a circular cross-section with radius $r$ in a spiral structure with two arms ($n_{\rm arms}=2$) provided by the following equation

\begin{equation}
    \dot{M}_{\rm H_2}=2m_{\rm p}N_{\rm H_2} v_{\rm in}\pi r^2 n_{\rm arms},
\end{equation}

where $m_{\rm p}$ is the proton mass and the inflow velocity  $v_{\rm in}=v_{\rm obs}/\sin{i}$, and $_{\rm obs}$ is the observed velocity inferred from the velocity residual maps (50\,\kms\ for NGC\,788 and NGC\,3516 and 80\,\kms\ for NGC\,5899), corrected by the inclination of the disc $i$ obtained from the modelling of the stellar velocity field by \citet{rogemar_stellar}.  The other term of this equation is $N_{\rm H_2}$, the molecular hydrogen density, obtained through $N_{\rm H_2}=\frac{M_{\rm H_2}}{2m_{\rm p}\pi r^2_{\rm d} h}$. $M_{\rm H_2}$ is the molecular gas mass as calculated by Eq.\,\ref{mh2} which is distributed in a thick disc with radius $r_{\rm d}$ with a height $h$ adopted as the typical value of $30$\,pc \citep{hicks09}. 

The masses of hot molecular gas range from 177 to 377\,M$_{\odot}$ and, as expected agree with the values we have already calculated for these galaxies \citep{astor19}. The densities, in the range $1.05-1.49\times10^{-3}$\cmm are systematically lower than we have previously measured in Mrk\,79 and in NGC\,2110: $3.3\times10^{-3}$\,\cmm \citep{rogemarM79} and $6.22\times10^{-3}$\,\cmm \citep{Diniz15}, respectively. Although we find this discrepancy the values have the same order of magnitude and can be considered consistent with each other.
In Table\,\ref{tab:inflow} the inflow properties are summarised. The inflow velocities are higher in NGC\,3516 reaching 160\,\kms. Mass inflow rates range from 0.1 to $0.9\times10^{-3}$\,M$_{\odot}$yr$^{-1}$ and are similar to values obtained for similar objects in the hot molecular gas \citep{rogemarM79,Diniz15,sb19}. 


\begin{table}
    \centering
    \caption{ Inflow properties: (1) Galaxy name; Physical scale, in parsec, of the (2) H$_2$ disc radius and of the (3) cross-section radius; (4) Velocity of the inflowing gas in \kms; (5) Mass inflow rate in 10$^{-3}$M$_{\odot}$yr$^{-1}$.}
    \begin{tabular}{lccccc}
    \hline
      Galaxy & $r_{\rm d}$ & $r$  & $v_{\rm in}$   & $\dot{M}_{\rm H_2}$ \\
       (1) & (2) & (3) & (4) & (5) \\
        \hline
        NGC788  & 278 & 139 & 140.8  & 0.9\\
        NGC3516 & 181 & 91 & 160.1  & 0.62 \\
        NGC5899 & 176 & 53 & 90  & 0.1\\
        \hline
    \end{tabular}
    \label{tab:inflow}
\end{table}

\subsection{Outflows Properties}
As mentioned in Sec.\ref{sec:summary}, in NGC\,5506, NGC\,3227 and NGC\,5899 the ionised gas shows signatures of outflows, traced by the broad component, 
while in Mrk\,607 an equatorial outflow is observed. The major part of this section is dedicated to estimate the mass, mass outflow rate and kinetic power of the ionised outflows. 

We also observe molecular outflows in NGC\,5899 and in Mrk\,607, traced by the distinct kinematic components. In the residual velocity maps (Sec.\ref{rotation}) evidence for the presence of molecular outflows are observed in NGC\,5506 and NGC\,3227. The hot molecular hydrogen represents a small fraction of the total molecular gas of the galaxy, thus we expect the mass outflow rates and powers to be small compared to the more powerful ionised outflows. The derived quantities will be presented and discussed in Sec.\,\ref{sec:out_mol}.

\subsubsection{Fraction of the gas in the outflows}
We assume the outflow is 
traced only by the broad component in NGC\,5506 and NGC\,3227. Thus, all the calculations performed in this section are based on this component. In NGC\,5899, both components fitted to the ionised gas emission lines are produced by a bipolar/biconical outflow and in Mrk\,607 the blueshifted and redshifted components are describing an equatorial outflow, thus we will adopt the flux of the outflowing gas as the sum of those of these two components.

The mass of H$^+$ in the outflow is estimated using the same equation presented in \citet{sbN4151Exc}, but instead of the flux of the \brg\ ($F_{\rm Br\gamma}$) we use the theoretical ratio between Pa$\beta$ and Br$\gamma$ fluxes for the case B recombination as $\frac{F_{\rm Pa\beta}}{F_{\rm Br\gamma}}=5.88$, assuming an electron temperature of 10\,000\,K and the low-density regime \citep{Osterbrock06} to obtain it in terms of the \pab\ flux, resulting in the following equation
\begin{equation}
 \left(\frac{M_{\rm H\,II}}{M_\odot}\right)=5.1\times10^{18}\left(\frac{F_{\rm Pa\beta}}{\rm erg\,cm^{-2}\,s^{-1}}\right)\left(\frac{D}{\rm Mpc}\right)^2\left(\frac{n_e}{\rm cm^{-3}}\right)^{-1},
\label{eq:mhii}
\end{equation}
where $F_{\rm Pa\beta}$ is the broad component \pab\ flux. We sum the fluxes over the spaxels within a radius of 100\,pc that corresponds to $\sim 1$\,arcsec or less, in these galaxies (see the scale in the continuum maps of Figs.\,\ref{fig:maps_N788}-\ref{fig:maps_M607}) and encloses the outflow structure in theses galaxies. We choose this same physical radius (100\,pc) for the outflow to allow the comparison between the galaxies. The electron density is given by $n_{e}$ and $D$ is the redshift distance of the galaxy. 

\citet{davies20} studied different methods to estimate the gas electron density in active galaxies in optical wavelengths.
The most traditional is the method based on the [S{\sc ii}]$\lambda6716,6731$ emission lines which provides typical densities for the NLR of $\sim500$\,\cmm \citep[e.g.][]{dors15,kakkad18,Freitas18}.
The method based on auroral [O{\sc ii}]$\lambda$7320,7331 and transauroral [S{\sc ii}]$\lambda$4069,4076 lines provides densities $\sim2000$\,\cmm, but its use is limited by the usual weakness of these emission lines which are hard to detect. The third method presented by the authors is based on the estimate of the ionisation parameter \citep{baron19} and systematically provides higher densities compared to those based on the [S\,{\sc ii}] doublet, reaching up to $22\,000$\,\cmm. High densities (10$^4$\,\cmm) are also observed using the  H-band [Fe\,{\sc ii}] \citep{sbN4151Exc,rogemar_cygnus} and optical [Ar\,{\sc iv}] \citep{rogemar_abundances} lines.  Our J and K band spectra do not include emission lines that can be used to determine the density of the outflows and to estimate the mass of the outflowing gas we have to assume a density value.  We follow the same procedure adopted by \citet{kakkad20} where two densities are used: one low (500\,\cmm) and one high (10\,000\,\cmm). 

The outflow masses, calculated inside the 100\,pc radius are presented in Table \ref{tab:my_masses} where $M\rm_{HII}^{ld}$ and $M\rm_{HII}^{hd}$ are the ionised hydrogen masses assuming 500\,\cmm and 10\,000\,\cmm densities, respectively. The masses estimates for the low-density regime are 20 times higher than the masses estimated assuming the higher density (which is the ratio between the gas densities).

For NGC\,5899, both components seen in the ionised gas emission lines are consistent with a bipolar outflow and the outflow dominates the emission within 100\,pc. For the other galaxies, we can compare the masses of the gas in the outflow with the total mass of ionised gas within 100\,pc, by computing the mass of gas in the disc using the fluxes of the narrow components. We find that the outflows correspond to fractions of 84\%, 46\% and 42\% for NGC\,3227, NGC\,5506 and Mrk\,607, respectively.


\subsubsection{Ionised gas mass outflow rates}
\label{sec:out_rates}
The geometry of the outflow impacts the way we estimate its rate and kinetic power \citep{lutz20}. Assuming the gas in the outflow is uniformly distributed inside a biconical or spherical geometry the mass outflow rate is given by \citep[e.g.][]{harrison14, fiore17, kakkad20}:

\begin{equation}
    \dot{M}_{\rm out}=3\frac{M_{\rm out}v_{\rm out}}{R_{\rm out}},
    \label{eq:out_rate}
\end{equation}
 where $M_{\rm out}$ is the mass of the gas that is outflowing, $v_{\rm out}$ is outflow velocity and $R_{\rm out}$ its radius. We consider $M_{\rm out}=M_{\rm HII}$ shown in Table~\ref{tab:my_masses} and calculated within $R_{\rm out} =100$\,pc. If we assume a biconical geometry, the outflow velocity ($v_{\rm out}$) is given by $\frac{v_{\rm obs}}{\sin{\theta}}$ with $v_{\rm obs}$, the observed velocity, inferred from the velocity fields and $\theta$ the inclination of the bicone with respect to the plane of the sky (see discussion below). On the other hand, if we assume the outflow is spherical we simply have $v_{\rm out}=v_{\rm obs}$. The observed velocity, fourth column of Table\,\ref{tab:my_masses}, is inferred from the \pab\ velocity fields of the components that represent the outflow (see the discussion above).
 
 \citet{fischer13} used long slit HST STIS data to observe and model the [O{\sc iii}]$\lambda5007$ as a biconical outflow in nearby galaxies, including  NGC\,3227 and NGC\,5006. In NGC\,3227 the outflow is almost perpendicular to the plane of the sky,  ($\theta=75^{\circ}$) in agreement with our broad component velocity fields (see Fig.\,\ref{fig:maps_N3227}), even though we do not observe the back part of the outflow as predicted by their model. For NGC\,5506 the biconical structure is not so clear in our data (see Fig.\,\ref{fig:maps_N5506}), but the 
 redshifts to the southeast of the nucleus contrasting with the blueshifts seen in the rest of the FoV in the \feii\ velocity field support its presence. In this case, the ionisation axis has a small inclination, $\theta=10^{\circ}$ \citep{fischer13}.  For Mrk\,607, from the blue and red components flux distributions and velocity fields (Fig\,\ref{fig:maps_M607}), we infer the outflow has a spherical symmetry. 
 
 The outflow in NGC\,5899 is not modelled by \citet{fischer13} and in \citet{ruschel-dutra21} an expanding spherical shape is assumed to reproduce the outflow and thus an inclination is not estimated. But the biconical structure is clear in the velocity fields of \pab\ and \feii\ of NGC\,5899. The zero velocity regions (see the central maps for both components of \feii\ and \pab\ in Fig\,\ref{fig:maps_N5899}) trace a a wall of the bicone parallel to the plane of the sky with the blueshifts and redshifts tracing the front and back walls, respectively. With this geometry in mind, we assume the inclination of the bicone is equal to the apparent opening angle, measured directly from the velocity fields, resulting in  $\theta=65^{\circ}$.



In Table\,\ref{tab:my_masses} we present the velocities and mass outflow rates for spherical (columns 4, 5 and 6) and biconical (columns 7, 8 and 9) geometries. The two different mass outflow rates for each geometry are obtained 
considering the low (superscript ld) and high-density (superscript hd) regimes (columns 2 and 3). In the case of Mrk\,607, the outflow is spherical, and we thus do not estimate its value for the biconical geometry.
Comparing the mass outflow rates for the two geometries we see that due to the high inclination assumed for NGC\,3227 both rates result
similar and we can thus infer the outflow is well represented by a spherical one. NGC\,5899 also does not have a big discrepancy between the two mass outflow rates. The biggest difference is seen for NGC\,5506, where the bicone axis is almost perpendicular to the plane of the sky leading to a high corrected velocity that leads to $\dot{M}_{\rm out}^{\rm ld}=12.5$\,M$_{\odot}$yr$^{-1}$ when we take into account this geometry. 

\subsubsection{Kinetic power}
\label{sec:out_power}
We can use the mass outflow rates derived in the previous section to estimate the power of the outflows by,
\begin{equation}
    \dot{E}_{\rm kin}\approx\frac{\dot{M}_{\rm out}}{2}\left(v^2_{\rm out}+3\sigma_{\rm out}^2\right).
\end{equation}
The term associated with $\sigma_{\rm out}^2$ is related with the disordered motions of the gas, which are significative in a gas in outflow. We adopt the lowest density and highest velocity regimes to assume $\dot{M}_{\rm out}$ as the value given in column 8 of Table\,\ref{tab:my_masses} and $v_{\rm out}=v_{\rm obs}/\sin{\theta}$. The velocity dispersion of the outflow, $\sigma_{\rm out}$ is estimated from the \pab\ broad component in NGC\,5506 and NGC\,3227, for the galaxies NGC\,5899 and Mrk\,607 we assume $\sigma_{\rm out}$ is the highest value from any of the components that represent the outflow. These values, as well as the $\dot{E}_{\rm kin}$ are presented in Table\,\ref{tab:my_masses}.



\begin{table*}
    \caption{Outflow properties: (1) Galaxy name; Masses of ionised gas calculated for (2) low density (500\,\cmm) and (3) high density (10\,000\,\cmm) in units of $\rm 10^4 M_{\sun}$; We assume spherical (columns (4)-(6)) and biconical geometries (columns (7)-(9)) with the outflow velocity is in units of \kms, and the mass outflow rates in $\rm M_{\sun}yr^{-1}$; (10) Velocity dispersion of the outflow in units of \kms; (11) Kinetic power, in units of $10^{40}$erg\,s$^{-1}$, calculated for the mass outflow rate in column (8), except in Mrk\,607 where the outflow is spherical and we use the value in column (5); (12) Kinetic efficiency of the outflows in percentage.
    }
    \centering
    \begin{tabular}{lcccccccccccc}
    \hline
   & & &\multicolumn{3}{c}{Sphere} &&\multicolumn{3}{c}{Bicone} &  & & \\
     \cline{4-6} \cline{8-10}\\
    Galaxy &  $M_{\rm HII}^{\rm ld}$ & $M_{\rm HII}^{\rm hd}$& $v_{\rm out}$ &   $\dot{M}_{\rm out}^{\rm ld}$ &  $\dot{M}_{\rm out}^{\rm hd}$ &&  $v_{\rm out}$&$\dot{M}_{\rm out}^{\rm ld}$ & $\dot{M}_{\rm out}^{\rm hd}$ & $\sigma_{\rm out}$& $\dot{E}_{\rm kin}$ & $\frac{\dot{E}_{\rm kin}}{L_{\rm bol}}$\\
    (1) & (2) &(3)&(4)&(5)&(6)&&(7)&(8)&(9)&(10)&(11)&(12)\\
    \hline
    NGC5506 & 70.90 & 3.54 &100&  2.17 & 0.11&& 576 & 12.49& 0.62 & 350& 275& 0.71\\
    NGC3227 & 3.72 & 0.19 &150 &0.17 & 0.008 && 155 & 0.18& 0.008 & 250& 1.18&0.02\\
    NGC5899 & 4.01 & 0.20 & 100&0.12 & 0.006 && 110 & 0.13 & 0.007 & 300& 1.20&0.03\\
    Mrk607  & 3.55  &0.18 & 100 & 0.108 & 0.005 && -  & - & - & 100& 0.13&0.005\\
    \hline
    \end{tabular}
    \label{tab:my_masses}
\end{table*}

\subsubsection{Molecular gas outflows}
\label{sec:out_mol}
In this section we determine the mass, mass outflow rate and power of the molecular outflows observed in the galaxies. In NGC\,5899 and Mrk\,607 the outflowing gas is assumed to be described by the broad component of the emission line.  The mass of the gas in the outflow can be determined by Eq.\,\ref{mh2}, where $F_{\rm H_2\,2.1218}$ is the flux of the broad component of the H$_2$ emission line. We determine 35\,M$_{\odot}$ and 26\,M$_{\odot}$ as the masses of the H$_2$ in the outflow for NGC\,5899 and Mrk\,607, respectively. 

A rough estimate of the mass outflow rate of the molecular gas can be obtained by $\dot{M}_{\rm out}^{\rm H2}=\dot{M}_{\rm out}\times\frac{M_{\rm H2}}{M_{\rm out}}$, where $M_{\rm out}$ and $\dot{M}_{\rm out}$ are the mass and outflow rate of the ionised gas outflow determined in Sec.\,\ref{sec:out_rates}. Thus, we determine the outflow rates of molecular gas being $1\times10^{-4}$\,M$_{\odot}$yr$^{-1}$ and $7.8\times10^{-5}$\,M$_{\odot}$yr$^{-1}$. For NGC\,5899, where the ionised gas seems to be distributed in a bicone, we find no difference in the mass outflow rate when taking into account this particular geometry. 

In the velocity residual maps (see Fig\,\ref{fig:rotation} and Sec.\,\ref{sec:rotation}) we observe residuals compatible with outflows in NGC\,5506 and NGC\,3227. Unlike the previous case, an isolated component is not describing this gas in the outflow, thus a different approach is necessary. We follow the procedure described in \citet{Diniz15} where a compact molecular outflow was observed in the residual velocity map for NGC\,2110. {
In NGC\,5506, signatures of an outflow weak outflow are observed and possibly associated with the outer layers of the more powerful ionised gas outflow} (see the third panel in the third column of Fig\,\ref{fig:rotation}), while in NGC\,3227 the structure is more compact, but the similar velocity intensities in each side make the assumption of a biconical geometry plausible. Using Eq.\,\ref{eq:out_rate} and assuming the biconical geometry, the molecular mass outflow rate is given by

\begin{equation}
  \dot{M}_{\rm out}^{\rm H_2}=2m_{\rm p}N_{\rm H_2}n v_{\rm out}\pi r_{\rm b}^2,
\end{equation}
where $m_{\rm p}$ is Tthe proton mass, $v_{\rm out}$ is the outflow velocity, $r_{\rm b}$ is the cross section radius of the bicone, 1\,arcsec for NGC\,5506 and 0.5\,arcsec for NGC\,3227, and $N_{\rm H_2}$ is the molecular hydrogen density estimated following the equation and assumptions in Sec.\,\ref{sec:inflow}. We calculate the density for a gas located in a disc with a radius $r_d=1$\,arcsec and a height $h=30$\,pc, 
resulting in $7\times10^{-3}$\,\cmm and $1.6\times10^{-2}$\cmm for NGC\,5506 and NGC\,3227, respectively.  

We use $n=2$ to account for both sides of the bicone. The observed velocity of the outflow is corrected by the inclination of the bicone with respect to the plane of the sky: $\theta=10^{\circ}$ for NGC\,5506 and $\theta=75^{\circ}$ for NGC\,3227 \citep{fischer13}. Thus, we calculate for NGC\,5506: $v_{\rm out}=143$\,\kms and $\dot{M}_{\rm out}^{\rm H_2}=5.5\times10^{-3}$\,M$_{\odot}$yr$^{-1}$; and for NGC\,3227: $v_{\rm out}=77$\,\kms and $\dot{M}_{\rm out}^{\rm H_2}=6.4\times10^{-4}$\,M$_{\odot}$yr$^{-1}$.  The kinetic power of the hot molecular outflows ranges from $10^{35}$ to $10^{38}$\,erg\,s$^{-1}$, being four orders of magnitude smaller than the powers estimated for the outflows of ionised gas. 

\section{The impact of inflows and outflows on the galaxies}
\label{sec:feed}
In order understand the impact of the observed inflows and outflows on the host galaxies, it is necessary to estimate the bolometric luminosity, $L_{\rm bol}$, of these objects. For the galaxies in our sample we can use the known Swift-BAT luminosity ($L_{\rm X}$) to estimate $L_{\rm bol}$ using $\log{L_{\rm bol}} = 0.0378(\log{L_{\rm X}})^2 - 2.03\log{L_{\rm X}}+61.6$ \citep{ichikawa17}. 

{
First, we can check if the inflows observed and quantified in NGC\,788, NGC\,3516 and NGC\,5899 can}
feed the AGN in each of these galaxies. In order to do so, we estimate the mass accretion rate by 

\begin{equation}
 \dot{m}=\frac{L_{\rm bol}}{c^2\eta},
\end{equation}
where is $c$ the light speed and $\eta$ is the rest frame mass conversion efficiency factor adopted as $\eta=0.1$ \citep{frank02}. We find $0.007\leq\dot{m}({\rm M_{\sun}yr^{-1}})\leq0.12$ which is higher than the mass inflow rate of hot molecular gas for any of these galaxies. This indicates that the inflows in hot molecular gas alone we observe are not enough to power the AGN and maintain their activity at the current luminosity. However, the hot molecular gas corresponds only to a small fraction of the total amount of gas in the inner region of galaxies, with masses of cold gas being 10$^5-$10$^7$ times larger than that in the hot molecular phase \citep[e.g.][]{dale05,ms06,mazzalay13}. 

The comparison between the kinetic power of the outflows and the bolometric luminosity, usually referred as the kinetic efficiency, is presented in the last column of Table \ref{tab:my_masses} for the outflows of ionised gas. \citet{hopkins_elvis10} found that for a hot gas outflow to be efficient in blowing the cold gas supply of the galaxy hosting an AGN its power needs to be, at least, 0.5\% of the galaxy's bolometric luminosity ($L_{\rm bol}$), which, in a direct comparison with our measurements, is only reached by NGC\,5506. {
However, \citet{harrison18} point out that the observed kinetic efficiencies, obtained from the gas kinematics, should not be directly compared to theoretical simulations. The observed kinetic efficiencies account only for the mechanical effect of the outflow seen in an specific gas phase and not the initial AGN energy input to the galaxy as usually presented in the models.}

We can also compare the mass accretion rate with mass outflow rate in NGC\,5506, NGC\,3227, NGC5899 and Mrk\,607. Specially NGC\,5506, the galaxy with the highest mass outflow rates (0.11-12.49\,M$_{\sun}$yr$^{-1}$ considering the different densities and geometries), has $\dot{m}=0.067$\,M$_{\sun}$yr$^{-1}$ which indicates more gas is disturbed and flowing out of the central region than it is directed to the feeding of the SMBH. This result agrees with the kinetic efficiency of $0.7\%$ we calculated previously. The only galaxy where we simultaneously observe inflows and outflows is NGC\,5899, which shows an outflow rate in ionised gas about one order of magnitude larger than the inflow rate in hot molecular gas. 

The molecular outflows we observe in the galaxies have a very low kinetic efficiency ($1.4\times10^{-6}-3.2\times10^{-5}$\, per cent), but, as stated before, the hot molecular gas represents a small fraction of the total mass of molecular gas and the kinetic efficiency does not account for all the impact an outflow can have on the host. Also, many assumptions are made in order to estimate the rates and powers and we understand they are highly uncertain. Nevertheless, we do observe disturbed molecular gas in four of our galaxies, indicating it not always displays ``well-behaved'' motions in the stellar disc. 

Thus, we find that the inflow rates in hot molecular gas in our sample are not enough to power their AGN activity at the current accretion rates (cold molecular gas may dominate the SMBH feeding processes) and the ionised outflows 
may only be able to redistribute the gas in the central kpc, which will be still available for future star formation.





\section{Conclusions}
\label{sec:conclusion}
We studied six nearby AGN hosts, from the AGNIFS selection of nearby active galaxies, namely NGC\,788, NGC\,3516, NGC\,5506, NGC\,3227, NGC\,5899 and Mrk\,607 using the Gemini NIFS integral field spectra in the J and K bands. For these galaxies we obtained the resolved ionised (\feiil\ and \pab) and molecular (\hml) gas distributions and kinematics in their inner kiloparsec, and investigated the AGN feeding and feedback processes using spatially resolved observations of inflows and outflows. Bellow, we summarise the main conclusions drawn from this work:

\begin{itemize}
    \item A multi-Gaussian fitting approach was adopted to describe the emission line profiles. The narrow Gaussian component traces the gas located in a disc that is dominated by a rotation pattern, except the ionised gas in NGC\,5899 that shows all of its emission coming from the outflowing gas. The broad -- and in one case the double red and blueshifted components -- trace non-circular motions. 
    
    \item A clear molecular outflow, traced by a broad component, is observed only for NGC\,5899. It is probably originated from the interaction between the ionised outflow with the molecular gas in the disc.
    \item Ionised gas outflows -- traced by the broad or double components -- are observed in four galaxies: NGC\,5506, NGC\,3227, NGC\,5899 and Mrk\,607.
    Molecular gas outflows are detected in NGC\,5899, interpreted as due to the interaction between the ionised outflow and the gas in the disc, and in Mrk\,607, interpreted as due to an equatorial outflow. 
    \item The molecular gas velocity field is usually well described by a rotating disc model, but the velocity residuals indicate the presence of gas inflows in three galaxies, NGC\,788, NGC\,3516 and NGC\,5899, and gas outflows in two, NGC\,5506 and NGC\,3227. Signatures of an equatorial outflow are observed in Mrk\,607. 
    
    \item The mass outflow rates of ionised gas in the four galaxies are in the range of $10^{-3}$ to $10^2$\,M$_{\odot}$yr$^{-1}$. For the molecular gas outflows we measure mass outflows rates in the range $10^{-5}$ to $10^{-3}$\,M$_{\odot}$yr$^{-1}$. 
    \item The hot molecular inflows cannot power and maintain the AGN at their current luminosity, as the mass inflow rates are three orders of magnitude lower than the mass accretion rates. This can be understood as due to the fact that the hot molecular gas traces only the ``hot skin'' of a much larger molecular gas reservoir. 
\end{itemize}

\section*{Acknowledgements}
We thank to an anonymous referee for the suggestions which helped us to improve this paper.
M.B. thanks the financial support from Coordenação de Aperfeiçoamento de Pessoal de Nível Superior - Brasil (CAPES) - Finance Code 001. R.A.R. acknowledges the support from Conselho Nacional de Desenvolvimento Cient\'ifico e Tecnol\'ogico  and Funda\c c\~ao de Amparo \`a pesquisa do Estado do Rio Grande do Sul.
R.R. thanks to Conselho Nacional de Desenvolvimento Cient\'{i}fico e Tecnol\'ogico  (CNPq, Proj. 311223/2020-6,  304927/2017-1 and  400352/2016-8), Funda\c{c}\~ao de amparo 'a pesquisa do Rio Grande do Sul (FAPERGS, Proj. 16/2551-0000251-7 and 19/1750-2), Coordena\c{c}\~ao de Aperfei\c{c}oamento de Pessoal de N\'{i}vel Superior (CAPES, Proj. 0001).
N.Z.D. acknowledges partial support from FONDECYT through project 3190769.
Based on observations obtained at the Gemini Observatory, which is operated by the Association of Universities for Research in Astronomy, Inc., under a cooperative agreement with the NSF on behalf of the Gemini partnership: the National Science Foundation (United States), National Research Council (Canada), CONICYT (Chile), Ministerio de Ciencia, Tecnolog\'{i}a e Innovaci\'{o}n Productiva (Argentina), Minist\'{e}rio da Ci\^{e}ncia, Tecnologia e Inova\c{c}\~{a}o (Brazil), and Korea Astronomy and Space Science Institute (Republic of Korea). 
This research has made use of NASA's Astrophysics Data System Bibliographic Services. This research has made use of the NASA/IPAC Extragalactic Database (NED), which is operated by the Jet Propulsion Laboratory, California Institute of Technology, under contract with the National Aeronautics and Space Administration.

\section*{Data Availability}
The data used in this work are publicly available online via the GEMINI archive https://archive.gemini.edu/searchform, under the following program codes: GN-2012B-Q-45, GN-2013A-Q-48, GN-2015A-Q-3, GN-2015B-Q-29 and GN-2016A-Q-6. The processed datacubes used in this article will be shared on reasonable request to the corresponding author.



\bibliographystyle{mnras}
\bibliography{refs} 




\appendix




\section{Channel maps}
\label{app:channel}
The channel maps of the \feiil, \pab\ and \hml for the six galaxies analysed here are presented as a supplementary file. They support the interpretation of the gas kinematics we present throughout the main body of the paper. 

\section{Notes on individual galaxies}
\label{app:notes_gals}
In this section, we discuss the results from Fig.\,\ref{fig:maps_N788}-\ref{fig:maps_M607} in details and put our measurements in context with previous results from the literature. All the distances to the galaxies are obtained from the redshift \citep{rogemar_sample}. The morphological classifications are from the RC3 catalogue \citep{devaucouleurs91} and the activity types are obtained directly from the Swift-BAT 105 month catalogue \citep{BAT105} unless stated otherwise. The names of the galaxies follow the same pattern of our previous works \citep{rogemar_stellar,rogemar_sample, astor19, rogemar_excitation}, but we indicate the alternative names in each subsection.

\subsection{NGC788}
This is a lenticular galaxy (SA0/a?(s)), with faint spiral arms seen in the HST F606W filter image \citep{martini03a}, at a distance of 57\,Mpc and hosting a Sy2 nucleus. In Fig.\,\ref{fig:maps_N788} we present the flux, velocity and velocity dispersion for \feii, \pab\ and H$_2$. The ionised gas distribution and kinematics in our maps agree with the ones shown in \citet{astor19}.
But due to the sky line superimposed on the spectrum (see Sec.\,\ref{sec:measurements}) we observe differences in the \hml\ velocity field -- ours show a larger velocity amplitude ($\sim 150$\,\kms) than theirs. The emission line flux distributions for the ionised and molecular gas are distinct. The \feii\ is most extended along the E-W direction with clumpy emission at the nucleus and at $0.8$\,arcsec east and west of it. The \pab\ flux distribution is more extended along the SW-NE direction, which is the approximate orientation 
of the galaxy minor axis \citep{rogemar_stellar}. The H$_2$ emission is distributed along the whole FoV with the highest fluxes observed along the N-S direction and extending to $0.5$\,arcsec from each side of the nucleus, almost perpendicular to the orientation of the \feii\ flux distribution.  This feature is compatible with the dusty molecular torus detected in CO(3-2) line emission, perpendicular to the ionised gas winds in Seyfert galaxies as observed in the Galaxy Activity, Torus and Outflows Survey  \citep[GATOS;][]{garcia-burillo21}. 
The velocity fields show evidence for a rotation pattern but heavily distorted. The velocity amplitude reaches 150\,\kms, which is twice the value observed for the stellar velocity field \citep{rogemar_stellar}. The lowest velocity dispersion values are observed for the H$_2$, where the smaller values are in the same regions as the low $\sigma_{\star}$ patches from \citet{rogemar_stellar}, and reaching values not higher than 80\,\kms. The highest values($\approx 130$\,\kms) are seen for the \feii, while the \pab\ has intermediate values of $\approx$100\,\kms. The higher velocity dispersion co-spatial with the highest fluxes for the \feii\ and distorted rotation pattern in the velocity field are indicatives of the presence of bipolar outflows. We further discuss these non-circular motions in Sec.\,\ref{sec:rotation}). 


\subsection{NGC3516}
NGC\,3516 is a lenticular galaxy at a distance of 37\,Mpc, classified as (R)SB0$^0$?(s), hosting a Sy1.2 nucleus, but recently the detection of a UV flare characterises this galaxy as a changing-look AGN \citep{ilic20}. Figure \ref{fig:maps_N3516} shows its flux distributions, velocity and velocity dispersion maps.
The peak of the emission in ionised and molecular gas is located at the nucleus and the emission-line flux distributions are similar to those obtained in \citet{astor19}. The \feii\ emission is more extended than that of \pab\ and is observed almost over the whole FoV. A 20\,cm continuum radio emission is observed to the north of the nucleus \citep{miyaji92}, but unlike \citet{barbosa09} that observe a correlation between the [S{\sc iii}]\,$9069$\AA\ emission and the radio, we do not see this feature in our ionised gas flux maps. The H$_2$ emission is extended over the whole FoV and is 
more extended to the SW of the nucleus which is consistent with the location of the dust lanes present there as reported by \citet{martini03a}.
 The velocity fields of ionised and molecular gas show rotation patterns. The highest velocity gradients are aligned in the NE-SW direction which is consistent with the orientation of the stellar \citep{barbosa06, rogemar_stellar} and [S{\sc iii}] \citep{barbosa09} velocity fields. The direction of blueshifts and redshifts is also consistent with those observed in [O{\sc iii}]$\lambda$5007 long-slit observations \citep{fischer13}. These similarities indicate that the gas may be located in the same plane as the stars. \pab\ and H$_2$ show the lowest velocity dispersion ($\sigma <150$\,\kms) with the smaller values observed for the former.
 The \feii\ presents higher velocity dispersion to the east of the nucleus consistent with the location of higher [S{\sc iii}] velocity dispersion (800\,\kms) where \citet{barbosa09} detected an excess of blueshifts associated with an ionised gas outflow. This scenario may also apply to the \feii\ as the velocity field deviates from the usual rotation pattern and the velocity dispersion is higher, reaching $\sim 200$\,\kms. The presence of a bipolar outflow was proposed by \citet{goad87} and more recently by \citet{fischer13} who reported high [O{\sc iii}] velocity dispersion values (600\,\kms) to the NE of the nucleus, at distances smaller than 1\,arcsec. Another hypothesis to explain the complex kinematics of NGC\,3516 is the presence of a twin-jet \citep{veilleux93} interacting with the ambient gas.

\subsection{NGC5506}
\label{app:n5506}
NGC\,5506, also known as Mrk\,1357, is a spiral galaxy classified as Sa pec edge-on, at a distance of 26\,Mpc and in a pair with NGC\,5507 \citep{tully15}. Its nucleus is classified as Sy1.9. 
In the top three rows of Fig.\,\ref{fig:maps_N5506} we present the flux, velocity and velocity dispersion of the narrow component of \feii, \pab\ and H$_2$ emission lines. 
The narrow component for the H$_2$ and \feii\ emission lines show the highest flux levels to the west and northwest of the nucleus which is consistent with the [O{\sc iii}]$\lambda5007$ emission from HST long-slit observations \citep{fischer13}. On the other hand, the narrow component of the \pab\ emission peaks $\sim 0.2$\,arcsec to the east of the nucleus. The molecular gas has a similar distribution to the one presented in \citet{astor19}. The same does not apply to the ionised gas: their maps are more similar to our broad component maps, as a consequence of their use of Gauss-Hermite functions being more sensitive to the wings in the emission line profiles and do not disentangle the multiple kinematic components of the line profiles. The velocity fields for the narrow components show mostly a rotation pattern, with a similar orientation to that observed for the stars \citep{rogemar_stellar}. Distortion from rotation pattern is observed mostly in the \feii\ velocity field (although some distortion is also seen for  \pab\ and H$_2$) that presents an excess of blueshifts ($\sim -50$\,\kms) to the north. 
Enhanced $\sigma$ values ($>150$\,\kms) are seen in the same region, indicating 
that this structure is likely tracing the emission of outflowing gas (see Sec.\ref{sec:rotation}). The outflow hypothesis is also supported by the fact that the stellar kinematics could not be mapped in the central arcsec because the CO absorption bands are diluted by the AGN radiation \citep{rogerio09, rogemar_stellar} and that a biconical ionised gas outflow perpendicular to the galaxy disc is detected by \citet{maiolino94}. 

The flux, velocity and velocity dispersion maps of the broad component of \feii\ and \pab\ are presented in the two bottom rows of Fig.\,\ref{fig:maps_N5506}. 
The fluxes for the broad components are up to one order of magnitude larger than those of the narrow components, indicating that most of the gas in NGC\,5506 is disturbed -- as the broader profiles are interpreted as due to such disturbances. As mentioned above, our broad component maps agree with those previously presented by \citet{astor19}. The velocity dispersion is larger than 300\,\kms over the whole FoV, indicating the gas is more disturbed most probably due to the presence of gas outflows. The velocity maps of the broad component are also indicative of the presence of outflows. The \pab\ velocity field shows redshifts over most of the FoV, except for a small blueshifted region to the north. For the \feii, velocities between zero and small negative values are seen at most locations, while some redshifts are observed to the south and southeast of the nucleus 
and some higher blueshifts are seen to the north. This pattern of velocity is consistent with the bi-conical outflow proposed by \citet{fischer13} to the north and south ($\rm PA=22^{\circ}$ based on the [O{\sc iii}] emission line. These authors suggest an opening angle of 40$^{\circ}$, but our velocity maps indicate this angle is larger possibly because the [O{\sc iii}] emission traces the higher ionisation that is more collimated.


In summary, for NGC\,5506 we find that the narrow components trace a rotating disc, but excess of blueshifts to the north associated with higher velocity dispersion indicates the presence of an ionised outflow. The broad components trace only the ionised outflow.

\subsection{NGC3227}
 NGC\,3227 is a spiral galaxy (SAB(s)a) interacting with the elliptical galaxy NGC\,3226 and forming the system Arp\,094 \citep{arp66}. Besides its companion, this galaxy resides in a group with 13 \citep{garcia93} or 14 \citep{crook07} galaxies, at a distance of 16\,Mpc and hosts a Sy1.5 nucleus. 
 The narrow component flux, velocity and velocity dispersion maps are presented in the top three rows of Fig.\,\ref{fig:maps_N3227} indicated by the green line on the right. The flux distributions of the \feii\ and H$_2$ are similar: both present the emission peak at $\sim0.3$\,arcsec southeast of the nucleus and an elongated structure approximately along the direction of the line of nodes of the stellar velocity field \citep[see Fig.\,\ref{fig:rotation} and ][]{rogemar_stellar}. The morphology of the H$_2$ emission structure within the inner 1\,arcsec$^2$ is in agreement with the SINFONI observations previously presented by \citet{davies06} at these scales. The \pab\ emission is more concentrated at the centre as already shown for the \brg\ emission \citep{davies06} and the total \pab\ emission \citep{astor19}, and shows a ring-like structure at 1\,arcsec from the nucleus. This circumnuclear ring has already been detected in cold molecular gas CO(2-1) \citep{schinnerer00, davies12}, with a radius of 1.5\,arcsec, and in the F160W HST image after the subtraction of a model for the bulge and disc emission \citep{davies06} in which the authors detect the ring at a radius of 1.7\,arcsec. Higher-resolution CO(2-1) (beam size of $0.214\times0.161$\,arcsec$^2$) ALMA observations unravelled the structure of the ring: several star forming clumps associated with star formation events \citep{alonso-herrero19}. The presence of a ring of star-forming regions is also supported by the low H$_2$/\brg\ ratios \citep{astor19, rogemar_agnifs}, higher \brg\ equivalent width
 \citep{rogemar_agnifs}, low gas ($\sim50$\,\kms\ - Fig.\,\ref{fig:maps_N3227}) and stellar velocity dispersion \citep{barbosa06,rogemar_stellar} values.
 The ionised and molecular gas velocity fields show amplitudes of 150\,\kms and signatures of a rotation pattern which, as for the other galaxies, is distorted indicating the presence of non-circular motions. The velocity gradient is orientted along the stellar line of the nodes (see Fig.\,\ref{fig:rotation}). Our H$_2$ velocity field agrees with that of the higher resolution (0.085\,arcsec) and smaller FoV ($0.8\times0.8$\,arcsec$^2$) K band SINFONI data obtained by \citet{davies06}, who proposed that velocity asymmetries along the galaxy minor axis can be caused by gas outflows. Recently, \citet{alonso-herrero19}, analysing the CO(2-1) kinematics, showed that streaming motions due to inflows associated with the large scale bar and circumnuclear star-forming ring, are the cause of the distortion in the molecular gas velocity field. These authors also interpret the CO(2-1) kinematics as being consistent with the presence of outflows at scales of $\sim 70$\,pc along the direction of the minor axis of the nuclear disc (PA\,$=50^{\circ}$).
The \pab\ and the \feii\ profiles have higher velocity dispersion values ($>100$\,\kms) distributed in an $0.5$\,arcsec wide arc-shaped structure inside the star-forming ring. This enhanced velocity dispersion structure suggests that a spherical shaped gas outflow might be present in NGC\,3227. The H$_2$ velocity dispersion is higher ($\sim 120$\,\kms) in a region perpendicular to the highest flux values and co-spatial with the distortion observed in the velocity field. \citet{davies06} interpret this feature as due to the fact that the H$_2$ is located in a thick disc instead of in a thin stellar disc.  This scenario is also supported by the ALMA data from \citet{alonso-herrero19} that interpret the nuclear emission of NGC\,3227 as due to molecular gas in a disc that is extended along tens of parsecs instead of in a compact torus.

On the two bottom rows of Fig\,\ref{fig:maps_N3227} we present the flux distribution, velocity and velocity dispersion for the \feii\ and \pab\ broad components.
The highest fluxes are somewhat extended to the north-northeast.
Also, as the broad component is overall brighter than the narrow component, the maps based on the fitting of the line profiles by a single Gauss-Hermite series shown in  \citet{astor19} are very similar to ours. One of the peaks of the radio emission in 18\,cm detected by \citet{mundell95} is offset $0.4$\,arcsec to the north of the nucleus,  which is co-spatial with the elongated structure in the  \feii\ flux distribution. Also, the highest \feii\ velocity dispersion values (250\,\kms) are seen at this same location
indicating that the gas is possibly being disturbed by the radio source. 
The velocity fields 
show only blueshifts, although the amplitudes are different: 150\,\kms\ for \pab\ and 250\,\kms\ for \feii. The velocity dispersion of both emission lines is higher than 200\,\kms with the highest values observed to the north of the nucleus for \feii. \citet{barbosa09} associate a compact radio emission at 3.6\,cm, also extended to the north, to the [S{\sc iii}]\,$9069$\AA\ outflows. The [S{\sc iii}] outflows are observed as blueshifts at PA\,$=-10^{\circ}$, which is similar to the orientation of our velocity fields, especially that for \feii. 

In summary, the narrow component in NGC\,3327 is mostly dominated by rotation, but signatures of non-circular motions are also present. The circumnuclear ring seen in the Pa$\beta$ emission is co-spatial with previous $^{12}$CO observations and seems to surround an spherical-bubble outflow. The broad component is tracing solely the ionised gas outflow.

\subsection{NGC5899}
NGC\,5899, at a distance of 36\,Mpc makes a pair with NGC\,5900 \citep{zaritsky97}. It is a spiral galaxy (SAB(rs)c) that hosts a Sy2 nucleus. 
In the top three rows of Fig.\,\ref{fig:maps_N5899} we present the maps for the narrow component, as indicated by the green line on the right. The H$_2$ flux distribution is elongated in the direction of the galaxy major axis (PA$=25^{\circ}$) and consistent with the flux map presented in \citet{astor19}. The flux distribution of the \pab\ and \feii\ are elongated along the north-south direction with two emission knots observed at $0.3$\,arcsec north and south of the nucleus. This orientation is consistent with the photometric major axis \citep{ruschel-dutra21}.
The H$_2$ velocity field shows a rotation pattern but its gradient seems to be slightly displaced from the orientation of the stellar line of the nodes (see Fig.\,\ref{fig:rotation}). The H$_2$ velocity dispersion shows overall low values ($<100$\,\kms) with the lowest values are seen to the north and the south of the nucleus and the highest in the perpendicular direction. Excess blueshifts, observed to the northwest of the nucleus, combined with the low velocity dispersion ($\sim60$\,\kms) of the region, may indicate the presence of streaming motions towards the centre of the galaxy (see Sec.\,\ref{sec:rotation}). Besides the distinct flux distributions, the velocity fields of  \pab\ and \feii\ also differ from that of H$_2$: blueshifts are observed mostly to the south and redshifts to the north, being opposite to the orientation of the molecular gas velocity field.
In addition, the velocity dispersion at the locations of the highest flux structures is larger than that of H$_2$. These differences indicate that the narrow components for the ionised gas are not originated from the gas in the disc, what seems to be the case for the molecular gas.  Some emission of gas located in the galactic plane is seen to the northeast, as indicated by the lower velocity dispersion values and similar velocity amplitudes for the ionised and molecular gas. 

The flux distributions, velocity and velocity dispersion for the broad component
are presented in the two bottom rows of Fig\,\ref{fig:maps_N5899}.
The flux maps of the H$_2$, \pab\ and \feii\ are similar: distributed along the north-south direction with the peak of the emission at $\approx 0.3$\,arcsec south of the nucleus. The emission is also more compact than for the narrow component. The H$_2$ velocity field show only blueshifts at $\sim-50$\,\kms and velocity dispersion of  $\sim250$\kms, which is lower than 
that of the broad components of the ionised gas emission lines. 
The \pab\ and \feii\ velocity fields for the broad component show mostly blueshifts, but some very low velocity redshifts ($\sim30$\,\kms) are seen to the north of the nucleus in the \feii\ velocity field. Their velocity dispersion maps present overall high values ($>250$\,\kms) with the highest ones observed to the north of the nucleus. This suggests that, as for the narrow component, the broad component in the ionised gas emission lines is not produced by gas in the galaxy disc, and both narrow and broad components trace an outflow that seems to have a bipolar geometry with its axis oriented along the north-south direction. The blueshifts probably trace the front wall of a cone-like outflow, the zero velocity spaxels the part that is aligned with the plane of the sky and redshifts the back of the cone.  These features lead us to suggest that the H$_2$ broad emission is produced by the interaction between the ionised outflow with the molecular gas 
in the disc.  

In summary, for NGC\,5899 the ionised gas kinematics seem to be dominated by a bipolar outflow, oriented approximately along the north-south direction. This outflow interacts with gas in the disc, pushing away molecular gas, which produces the broad H$_2$ component. Most of the molecular gas seems to be located in the disc of the galaxy.


\subsection{Mrk607}
Mrk\,607, also known as NGC\,1320 and MCG-01-09-036, is an edge-on spiral galaxy with inclination $i=70^{\circ}$ \citep{rogemar_stellar}, classified as Sa?, hosts a Seyfert 2 nucleus and is at a distance of $\approx 37$\,Mpc. It is a member of a small group of galaxies, together with NGC\,1321 and NGC\,1289 \citep{crook07}.
In the top sub-figure of Fig\,\ref{fig:maps_M607} we present the flux distribution, velocity and velocity dispersion for what we call the disc component 
The flux distributions for the molecular and ionised gas are elongated along the NW-SE direction, approximately the direction of the galaxy major axis, PA$=138^{\circ}$ \citep{rogemar_stellar}. The highest flux values are observed at and near the nucleus for the ionised gas, but for the H$_2$ the flux distribution is more homogeneous over the whole FoV. Similar extensions and orientations are observed in the flux distributions of optical emission lines, for instance [O{\sc iii}]$\lambda5007$, H$\alpha$, [N{\sc ii}]$\lambda6583$ and [S{\sc ii}]$\lambda6731$ \citep{Freitas18}. Our flux distributions are also similar to the HST [O{\sc iii}] flux distribution \citep{schmitt03} where the strongest emission is observed at the nucleus and extended to up to 3.75\,arcsec to the northwest, in the direction of the galaxy major axis, which is consistent with the \pab\ flux distribution. The velocity fields present a well-defined rotation pattern, with a velocity amplitude of $\sim200$\,\kms. 
Although the gradient of the stellar velocity field \citep{rogemar_stellar} is oriented in the same direction, the gas is rotating in the opposite direction. This result was also previously reported by \citet{Freitas18}. The counter-rotation between gas and stars suggests that these components have different origins. Since the stellar disc is more stable than the the gas, and thus the orbits of stars are harder to change, it is likely that the gas kinematics have been affected by the interaction between Mrk\,607 and its closest companion, NGC\,1321.

In the two sets of panels in the bottom of Fig.\,\ref{fig:maps_M607} we present the maps for the blueshift component, to the left, and for the redshift component, to the right. As the region fitted with three Gaussian functions corresponds only to the central arcsecond or so we zoomed in the central $1\times1$\,arcsec$^2$ region of the FoV shown as the green square in the  H$_2$ flux map of the disc component.  
A similar 
kinematic pattern (with one blueshifted and one redshifted component) around the nucleus has been observed in other Seyfert galaxies, e.g. NGC\,5929 \citep{rogemar_n5929}, interpreted as being produced by an equatorial outflow, i.e. the torus itself expanding 
outwards. This interpretation is proposed because the highest velocity dispersion values -- or the two components here, are observed perpendicular to the observed radio emission axis.
Previous optical IFS of Mrk\,607 also shows an enhancement in the gas velocity dispersion perpendicularly to the galaxy disc \citep{Freitas18}, while the radio continuum emission is extended along the north-south direction, with some faint emission also to the north-west along the major axis \citep{colbert96}. The beam size of this observation is  given by a Gaussian with FWHM of 3\,arcsec, the size of our field of view, thus a clear association between the jet orientation and the outflow cannot be made. 
The peak of the flux distributions of the blueshifted and redshifted components are slightly displaced from the nucleus: the blueshifted component emission peaks northwest of the nucleus, while the redshifted emission peaks southeast of the nucleus.

In summary, for Mrk\,607, both the ionised and molecular gas present two kinematic components: (i) one due to emission of gas rotating in the galaxy disc, in the opposite direction of the stellar motions and (ii) an equatorial outflow or compressed gas by wide-opening angle outflows in colder gas phases, identified by the blueshifted and redshifted emission-line components.
\bsp	
\label{lastpage}
\end{document}